\documentclass[fleqn,usenatbib]{mnras}

\usepackage{lineno}
\usepackage{newtxtext,newtxmath}
\usepackage{multirow}
\usepackage[T1]{fontenc}

\DeclareRobustCommand{\VAN}[3]{#2}
\let\VANthebibliography\thebibliography
\def\thebibliography{\DeclareRobustCommand{\VAN}[3]{##3}\VANthebibliography}

\usepackage{graphicx}	
\usepackage{amsmath}	

\title[Radial Sizes of CMEs at Different Instances]{Non-conventional Approach for Deriving the Radial Sizes of Coronal Mass Ejections at Different Instances: Discrepancies in the Estimates Between Remote and In Situ Observations}

\author[Agarwal and Mishra]{
Anjali Agarwal$^{1,2}$\thanks{E-mail: anjali.agarwal@iiap.res.in}
and Wageesh Mishra$^{1}$\thanks{E-mail: wageesh.mishra@iiap.res.in}
\\
$^{1}$Indian Institute of Astrophysics, II Block, Koramangala, Bengaluru 560034, India\\
$^{2}$Pondicherry University, R.V. Nagar, Kalapet 605014, Puducherry, India\\
}

\date{Accepted XXX. Received YYY; in original form ZZZ}

\pubyear{2024}

\begin{document}
\label{firstpage}
\pagerange{\pageref{firstpage}--\pageref{lastpage}}
\maketitle


\begin{abstract}
Understanding the evolution of radial sizes and instantaneous expansion speeds of coronal mass ejections (CMEs) is crucial for assessing their impact duration on Earth's environment. We introduce a non-conventional approach to derive the CME's radial sizes and expansion speeds at different instances during its passage over a single-point in situ spacecraft. We also estimate the CME's radial sizes and expansion speeds during its journey from the Sun to 1 AU using the 3D kinematics of different CME features, including the leading edge (LE), center, and trailing edge (TE). The continuous 3D kinematics of the CME is estimated by employing the GCS and SSSE reconstruction methods on multi-point observations from coronagraphs and heliospheric imagers combined with the drag-based model. We choose the 2010 April 3 CME as a suitable case for our study, promising a more accurate comparison of its remote and in situ observations. We show that the introduced non-conventional approach can provide better accuracy in estimating radial sizes and instantaneous expansion speeds of CMEs at different instances. We examine the aspect ratio of the CME, which influences its expansion behavior and shows the discrepancy between its value in the corona and interplanetary medium. Our study highlights significant inconsistencies in the arrival time, radial size, and expansion speed estimates obtained from remote and in situ observations. We advocate for future studies leveraging multi-spacecraft in situ observations and our non-conventional approach to analyze them to improve the comprehension of CME dynamics in the solar wind.
\end{abstract}

\begin{keywords}
Sun: coronal mass ejections (CMEs) -- Sun: corona -- Solar-terrestrial relations
\end{keywords}



\section{Introduction}

Coronal mass ejections (CMEs) are the huge expulsions of the magnetized plasma bubble from the Sun into the heliosphere and are the primary drivers of adverse space weather effects \citep{Schwenn2006,Pulkkinen2007,Webb2012,Schrijver2015}. CMEs are often remotely observed in white light, using coronagraphs (CORs) and heliospheric imagers (HIs), due to Thomson scattering of photospheric light by the electrons in the solar corona \citep{Billings1966,Howard2009,Howard2013}. CMEs can also be observed in in situ observations that can provide measurements of CME parameters along a 1D cut made by the in situ spacecraft through the CME \citep{Burlaga1981,Crooker1996,Bothmer1998}. There have been several attempts to use observations from multiple spacecraft, combined with modeling efforts, to understand the 3D kinematics, global morphology, radial size, and propagation behavior of CMEs \citep{Sheeley1999,Xie2004,Jian2008b,Lugaz2010,Nieves-Chinchilla2013,Winslow2015,Kilpua2019,Lugaz2020b,Mishra2021,Zhuang2023}. 
In the current era, heliospheric imaging and in situ observations from unprecedented locations closer to the Sun have made good progress in understanding the evolution of CMEs in the preconditioned ambient medium \citep{Davies2009,Mishra2017,Möstl2022,Khuntia2023,Berriot2024,Palmerio2024}. However, due to difficulty in unambiguously tracking features at distances far from the Sun and further identifying them in in situ observations, it is still challenging to establish the one-to-one association among the features observed in these two sets of observations \citep{Kilpua2017,Mishra2023,Temmer2024}.

Predicting the arrival time of CMEs or magnetic clouds (MCs) at the Earth is important \citep{Webb2012,Vourlidas2019,Temmer2024} for the onset of space weather phenomena while the radial size, impact duration, momentum, and magnetic field can govern the intensity of the perturbations and recovery time for the disturbed magnetosphere to restore its quiet state \citep{Gonzalez1999,Wang2003,Srivastava2004,Echer2008,Wood2017}. Despite the importance of radial sizes of CMEs on the Earth, only a handful of studies have investigated the continuous evolution of radial sizes during the heliospheric journey of CMEs \citep{Savani2009,Wood2017}. The radial dimension of a CME is expected to be linked to its radial expansion. A faster propagating CME has a larger expansion speed and consequently can be of a bigger size \citep{Owens2005}. Although several attempts have been made to estimate the lateral expansion speed of CMEs closer to the Sun using coronagraphic observations and connecting it to the radial propagation speed of CMEs \citep{Schwenn2005,Gopalswamy2009,Scolini2019,Balmaceda2020}, only limited studies are reported to derive the radial expansion speeds from such observations \citep{Savani2009,Patsourakos2010}. Understanding the evolution of radial sizes of CMEs can help us better understand the physical processes governing the expansion, a relative decrease in the thermal and magnetic pressure content inside the CMEs and solar wind, and the increasing separation between different features/substructures (leading edge, center, and trailing edge) of the CMEs.

There have been several attempts to estimate the radial sizes of CMEs at different distances from the Sun. Using multi-spacecraft (Voyager 1 \& 2, Helios 1 \& 2, and IMP 8) in situ observations, \citet{Burlaga1981} estimated the radial size of MCs. In another attempt, using multi-spacecraft (IMP, Pioneer 11, and Pioneer 10) in situ observations, \citet{Crooker1996} found that MCs can have highly distended cross sections, with longitudinal dimension exceeding radial dimension by at least a factor of 8. These studies are based on analyzing a few selected cases measured in situ and do not provide a connection to the estimates of CME characteristics derived from imaging observations. There are also statistical studies estimating the radial sizes of CMEs on the Earth over different solar cycles \citep{Zhang2008,Mitsakou2014,Kilpua2017,Mishra2021a}. Notably, the local in situ measurements cannot differentiate if the measured characteristics of the CME are inherent or due to the evolution of CME in the surrounding medium or if it is merely the effect of spacecraft trajectory through the CME. Further, such studies pose limitations as detecting the same feature of a CME at multiple spacecraft, which are often not radially aligned, is rarely possible.

There are interesting studies, but only a handful, combining multi-viewpoint remote observations and in situ measurements from radially aligned spacecraft to investigate the radial dimension and expansion of CMEs up to 1 AU \citep{Nieves-Chinchilla2012,Nieves-Chinchilla2013,Lugaz2020b}. Earlier studies have often focused on tracking a CME bright leading edge in the imaging observations (coronagraphs or HIs), with only a few tracking the cavity or filament of a CME to derive their 3D kinematics \citep{Liu2010,DeForest2011,Colaninno2013,Mishra2013,Möstl2013,Mishra2014a,Temmer2014,Mishra2015,Rouillard2020}. Also, several models (empirical, analytical, and MHD) have mostly attempted to investigate the evolution of the CME leading edge \citep{Gopalswamy2000,Odstrcil2004,Schwenn2005,Vršnak2010,Scolini2019,Mayank2024}. Investigating the evolution of different substructures (leading edge, center, and trailing edge) of a CME can provide a better understanding of the relative forces acting on them, the evolution of their propagation and expansion speeds, and their radial dimensions.

Progress toward accurately estimating the evolution of radial sizes of CMEs will be crucial for estimating the expansion speeds, arrival times of various substructures, and the longevity of space weather events. Given the limited number of studies focusing on the ongoing changes in the radial size and expansion speed of CMEs \citep{Savani2009,Nieves-Chinchilla2012}, it is imperative to analyze more cases and interpret the findings in the context of earlier studies. Additionally, different substructures, such as the leading edge (LE), center, and trailing edge (TE), of CMEs/MCs may exhibit different characteristics in different instances \citep{DeForest2011,Mishra2015}. However, most existing studies utilizing remote observations and modeling primarily examine only the arrival time of the CME leading edge and compare it with in situ observations \citep{Liu2010,Mishra2013,Scolini2019,Mayank2024}. Furthermore, the conventional approach estimates the radial expansion speed of CMEs as half of the difference between the leading edge and the trailing edge speed measured at a certain location of the spacecraft \citep{Crooker1996,Owens2005,Jian2008b,Richardson2010}. Since the arrival of the leading edge and trailing edge of the CME at 1 AU are often separated by several hours, the conventional approach cannot accurately provide the instantaneous expansion speed at the arrival of any CME substructures (LE, center, and TE) at a certain in situ spacecraft.

This paper focuses on a non-conventional analysis approach to single-point in situ observations to estimate the CME's radial size and instantaneous expansion speed. We describe our non-conventional approach as considering different accelerations of each substructure of the CME, which implies that the CME has a non-zero constant expansion acceleration during its passage at the in situ spacecraft. These estimates from the non-conventional approach are compared with the radial size and expansion speeds derived from multi-point remote observations combined with the drag-based model. To demonstrate the concept of our study, we selected a CME of 2010 April 3. Earlier studies have focused on estimating the arrival time and propagation speed of only the leading edge (LE) of this CME \citep{Liu2011,Mishra2013,Colaninno2013,Mishra2014a,Wood2017}. Also, studies have undertaken the geo-effectiveness of this CME and its shock \citep{Möstl2010,Xie2012,Hess2017}. Different models have been implemented to examine the kinematics and thermodynamic properties of the CME \citep{Wood2011,Mishra2020}. It is evident that the 2010 April 3 CME has been studied extensively in the literature from different perspectives, but none of the earlier studies have focused on examining the evolution of radial sizes.

Selecting the extensively studied CME of 2010 April 3 offers numerous advantages: (i) Since our approach is to estimate radial sizes of the CME utilizing its 3D kinematics of different substructures at varying distances from the Sun, we can validate our estimates of the CME LE kinematics against previously established studies. (ii) Earlier studies focused on estimating the kinematics of this CME LE only; our estimates of 3D kinematics for both the center and TE will enhance our comprehension of the distinct kinematics exhibited by various substructures of the CME. (iii) The chosen CME, characterized by high speed and minimal deceleration beyond coronagraphic heights, promises to be an ideal candidate for a reasonable comparison between remote and in situ observations. This implies the possibility of minimal inconsistency in the estimates from the remote and in situ observations to demonstrate one of the best performances of the conventional methods in forecasting the radial size, expansion speeds, and impact duration of CMEs at 1 AU. (iv) This non-decelerating CME could also be a good candidate for validating the kinematics obtained from methods applied to HIs observations by comparing it to those obtained from coronagraphic observations, given more challenges in reliably tracking CMEs and estimating their kinematics away from the Sun.

The utilized in situ observations and their analysis using our non-conventional approach are described in Section~\ref{sec:insitu}. The kinematic evolution of different substructures of the CME as obtained from remote observations is described in Section~\ref{sec:rem}. The inconsistencies in the estimates from both sets of observations are noted in Section~\ref{sec:com}. The role of the aspect ratio of the CME in governing the radial size and expansion speeds of the CME is outlined in Section~\ref{sec:aspect}. Section \ref{sec:resdis} summarises our results and discusses the factors that can bring some uncertainties in our findings.

\section{Observations of Selected CME and Analysis Methodology} \label{sec:obsana}

We investigate the evolution of radial sizes of the selected 2010 April 3 CME at different instances using in situ and remote observations. In this work, we utilized the in situ observations of the CME from Wind spacecraft near 1 AU \citep{ Ogilvie1997} and identified the CME boundaries based on the magnetic field and plasma parameters described in Section~\ref{sec:insitu}. We focus on our non-conventional approach to the in situ observations from single-point spacecraft to estimate the radial sizes and instantaneous expansion speeds as described in Section~\ref{sec:app}. For investigating the continuous evolution of the CME, we use its remote observations from white light coronagraphs \textit{Large Angle and Spectrometric COronagraph} (LASCO) onboard \textit{SOlar and Heliospheric Observatory} (SOHO), coronagraphs (CORs) and heliospheric imagers (HIs) observations onboard twin \textit{Solar TErrestrial RElations Observatory} (STEREO) spacecraft \citep{Brueckner1995,Kaiser2008,Howard2008,Eyles2009}.

In the following, we first describe the in situ observations-based estimates of CME radial sizes and speeds at different instances when LE, center, and TE of the CME arrive at 1 AU. Thereafter, we describe the remote observations of CMEs to estimate the 3D kinematics of the CME using conventional 3D reconstruction methods. The estimated 3D kinematics of different substructures (LE, center, and TE) of the CME are used to derive the radial size and instantaneous expansion speed of the CME. The derived radial size and instantaneous expansion speed from remote observations are compared with the estimates from in situ measurements at 1 AU. Our analysis focuses on the inconsistencies in the estimates from both sets of observations and investigates the possible reasons for the same.

\subsection{In Situ Observations of the CME}\label{sec:insitu}

\begin{figure}
    \centering
    \includegraphics[scale=0.6]{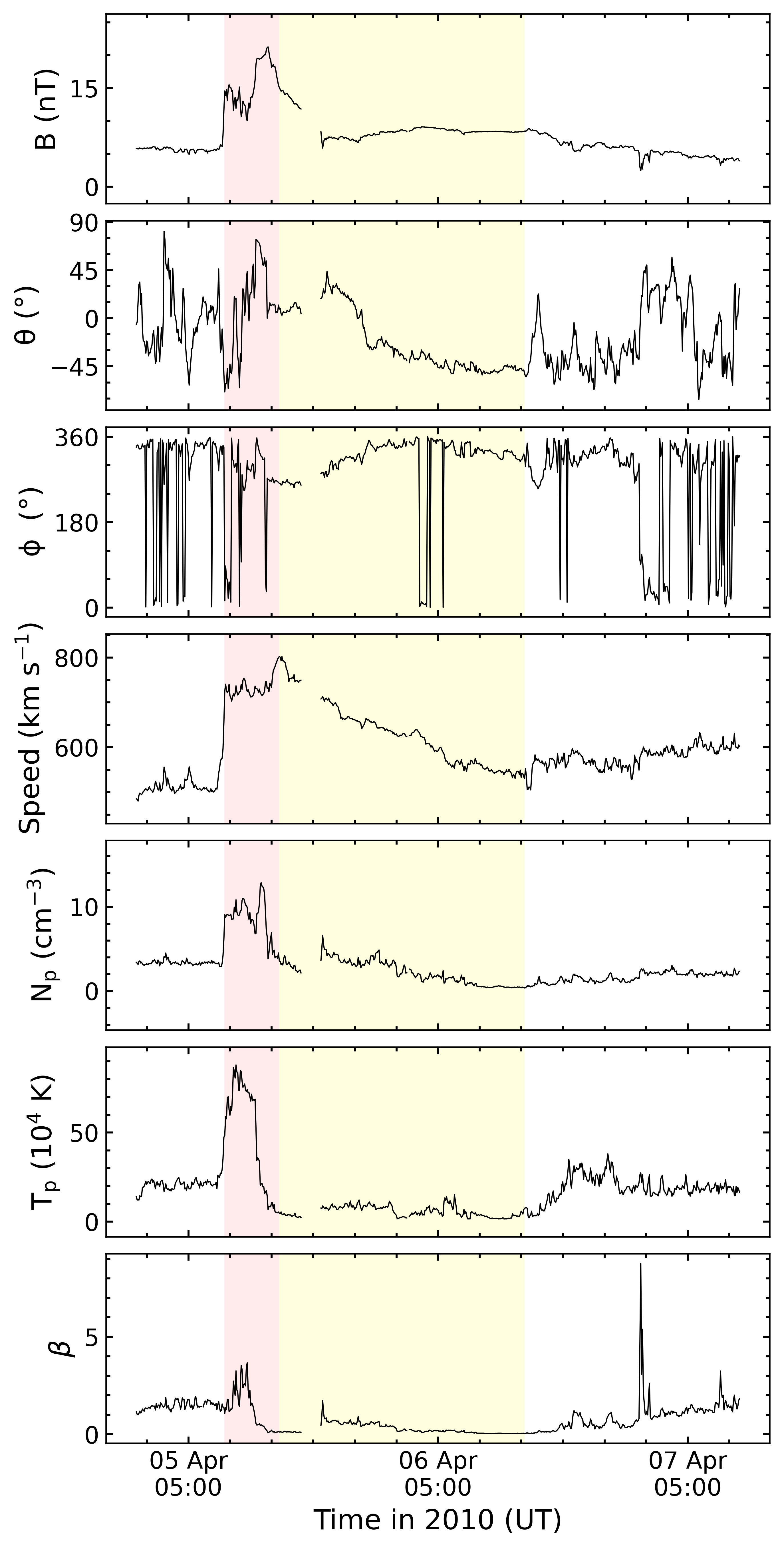}
    \caption{The top to bottom panels show the variation of the magnitude of the magnetic field, latitude and longitude of magnetic field vector, proton speed, proton density, proton temperature, and plasma beta. Transparent fill areas with red and yellow represent the sheath and magnetic cloud duration during the passage of the CME on in situ spacecraft at 1 AU.}
    \label{fig:insitu}   
\end{figure}

For the in situ observations of the 2010 April 3 CME, we employ data from the Wind spacecraft in GSE coordinate located close to 1 AU (at the L1 point). Figure~\ref{fig:insitu} illustrates the in situ observation of the CME; the panels from top to bottom show the magnetic field magnitude, $\theta$, $\phi$, speed, proton density, proton temperature, and plasma beta. We estimate $\theta$ using the magnitude of the total magnetic field and normal component of magnetic field (${B_z}$) as ${\theta = \sin^{-1}(\frac{B_z}{B})}$. Since $\phi$ rotate in the ecliptic plane (from 0$^\circ$ to 360$^\circ$), therefore, $\phi$ is estimated using magnetic field components ${B_x}$ and ${B_y}$ as for ${B_x>0}~\text{and}~{B_y>0}$, ${\phi = \tan^{-1}(\frac{B_y}{B_x})}$; for ${B_x < 0}~\text{and}~{B_y > 0}$, ${\phi = \tan^{-1}(\frac{B_y}{B_x})} + {180^\circ}$; for ${B_x < 0}~\text{and}~{B_y < 0}$, ${\phi = \tan^{-1}(\frac{B_y}{B_x})} + {180^\circ}$; for ${B_x > 0}~\text{and}~{B_y < 0}$, ${\phi = \tan^{-1}(\frac{B_y}{B_x})} + {360^\circ}$.

We scrutinize in situ data to discern the boundaries of CME or magnetic cloud (MC), employing multiple signatures simultaneously as described by \citet{Zurbuchen2006}. Figure~\ref{fig:insitu} illustrates the arrival of the CME shock at 08:28 UT on April 5, with the sheath duration of the CME depicted by a fill area in transparent red. The CME/MC LE reaches Wind at 13:43 UT on April 5, while its TE reaches Wind at 13:20 UT on April 6. Consequently, the transparent yellow fill area denotes the 23.6 hours of duration of the MC. Our estimates of the MC boundaries are in good agreement, within 1 hour, with several earlier studies \citep{Möstl2010,Liu2011,Mishra2013,Mishra2014a}. This figure explores the rotation of $\theta$ and $\phi$ within the MC boundary to observe cloud orientation, which is the North-West-South (NWS) direction. In the fourth panel, a linear decrease in speed between the MC boundary signifies the expansion of the cloud.

Using the conventional approach \citep{Owens2005}, we used the in situ measured speeds of the LE and TE of 2010 April 3 MC, and estimate the expansion speed ($V_{exp}$ = $\frac{V_L - V_T}{2}$) of the MC as 136.5 km s$^{-1}$. The conventional approach assumes a constant expansion speed throughout the entire passage of the MC over a specific in situ spacecraft. Consequently, the calculated expansion speed is not instantaneous at different instances corresponding to the arrival of different features of the MC. In the following Section~\ref{sec:app}, we outline our non-conventional approach, which enables us to estimate the radius of the MC at distinct instances (at the arrival of LE, center, and TE) and determine the instantaneous expansion speed from single-point in situ measurements.

\subsubsection{Non-conventional Approach to Examine the Evolution of Radial Size and Expansion Speed from Single-point In Situ Observations} \label{sec:app}

\begin{figure}
    \centering
    \includegraphics[scale = 0.32,trim={0.2cm 0cm 0cm 0cm},clip]{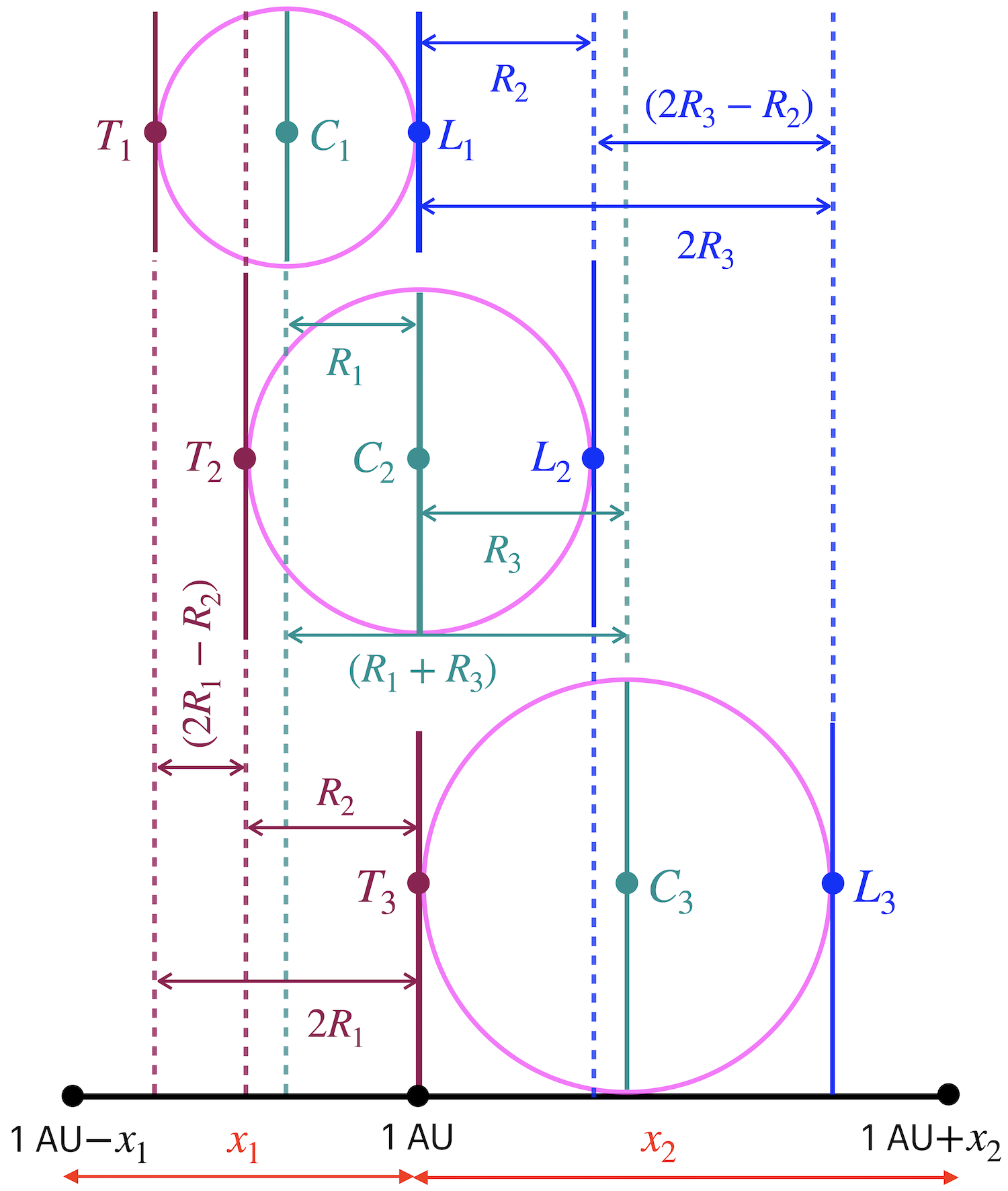}
    \caption{The schematic illustrates the evolution of an expanding CME during its passage over the in situ spacecraft. The magenta circles represent the geometry of a CME in the plane of an in situ spacecraft. The blue, green, and maroon vertical lines denote the LE, size center, and TE of the MC, respectively. The location of in situ spacecraft at 1 AU is marked on the horizontal black line with two additional distances, one greater than 1 AU (1 AU + $x_2$) and one lesser than 1 AU (1 AU - $x_1$). The top to bottom panels represent the arrival of LE (L), center (C), and TE (T) at 1 AU at different instances of $t_1$, $t_2$, and $t_3$, respectively. The left-right arrow represents the distance traveled by different features and the evolution of the CME radial dimension during any two instances.}
    \label{fig:MC_1Dcut}
\end{figure}

We demonstrate a non-conventional approach to examine the evolution of radial size and instantaneous expansion speed at different instances during the passage of the MC over the in situ spacecraft. The conceptual representation of this approach is illustrated in Figure~\ref{fig:MC_1Dcut}. Irrespective of the assumed CME structure, the in situ spacecraft will provide 1D measurements of the CME plasma parameters along its trajectory through the CME. Even in the case of a flank encounter of the CME with the in situ spacecraft (i.e., without the intersection of the spacecraft along the nose of the CME), the sampled region of the CME can be classified into leading, center, and trailing portions. For simplicity of explaining our non-conventional approach, we can assume the CME flux rope as a circle in the plane of spacecraft (the orbital plane of the spacecraft, which is the ecliptic plane in our study as we are using in situ observations of WIND at L1).

For the schematic representation, the magenta-colored circle in the figure shows the MC in the ecliptic plane. The circular MC shown in the top, middle, and bottom panels shows the arrival of LE, center, and TE at 1 AU at the instances of $t_1$, $t_2$, and $t_3$, respectively. The blue, green, and maroon vertical solid lines touching/intersecting the circle denote the LE, center, and TE of the MC, respectively. The LE, center, and TE of the MC are denoted with symbols $L$, $C$, and $T$, respectively, and those with subscripts 1, 2, and 3 represent the scenario at three different instances $t_1$, $t_2$, and $t_3$. The location of the in situ spacecraft at 1 AU is indicated on the horizontal black line. The marking of 1 AU distance on the horizontal line is clear with two additional distances, one greater than 1 AU (1 AU + $x_2$) and another lesser than 1 AU (1 AU - $x_1$). The radius of the cloud at the arrival of LE, center, and TE at 1 AU (i.e., at $t_1$, $t_2$, and $t_3$) is $R_1$, $R_2$, and $R_3$, respectively. The increasingly bigger size of the circular MC at $t_1$, $t_2$, and $t_3$ denotes the expansion of the MC. The figure also shows the distance traveled by each feature (L, C, and T) between any two instances, such as $t_1$ to $t_2$, $t_2$ to $t_3$, and $t_1$ to $t_3$. The center considered in this approach is the size center that divides the in situ sampled radial size of the MC equally into two parts. In the in situ observations, one can also mark the time center, which equally divides the MC's duration into two parts. In Section~\ref{sec:insitu_estimate}, we would describe the situation where the arrival of the time center and size center of MCs at in situ spacecraft can differ. Hereafter, we refer to the size center as the center.

In contrast to the conventional approach, the instantaneous (at $t_1$, $t_2$, and $t_3$) expansion speed of the MC, assuming its constant acceleration, can be estimated using the propagation speed of different features at the same instance (any of $t_1$, $t_2$, and $t_3$). However, the single-point in situ measurements of the MC provide the propagation speeds of different features at different instances. In our approach, we estimate the speeds of a particular feature (any of $L$, $C$, and $T$) at different instances using the in situ measured speed of that particular feature at a single instance. For this purpose, we use the first equation of motion as:


$${V_{F}({t_j}) = V_{F}({t_i}) + a_{F}{t_{ji}}}$$

where the subscript $F$ stands for features (any of $L$, $C$, and $T$) and duration $t_{ji}$ is the difference between two instances as $t_j - t_i$ (for $j > i$). Thus, $V_{F}({t_j})$ and $V_{F}({t_i})$ denotes the speed of features at time $t_j$ and at time $t_i$ and $a_{F}$ is the constant acceleration of the respective feature during the passage of MC at in situ spacecraft. The estimated speeds of different features at the same instance can be used to calculate the instantaneous expansion speed of the MC.

Further, using the demonstrated non-conventional approach, we estimate the radial size of the MC at different instances and the distance traveled by different features between any two instances (as labeled in Figure~\ref{fig:MC_1Dcut}). The distances traveled by LE, center, and TE from $t_1$ to $t_2$ are $R_2$, $R_1$, and $2R_1 - R_2$, respectively. These traveled distances can be expressed using the second equation of motion as:

$$R_2 = V_L(t_1)t_{21} + \frac{1}{2}a_Lt_{21}^2$$ 
$$R_1 = V_C(t_1)t_{21} + \frac{1}{2}a_Ct_{21}^2$$ 
$$2R_1 - R_2 = V_T(t_1)t_{21} + \frac{1}{2}a_T t_{21}^2$$

Similarly, the distances traveled by LE, center, and TE from $t_2$ to $t_3$ are $2R_3 - R_2$, $R_3$, and $R_2$, respectively. Further, the distances traveled by LE, center, and TE from $t_1$ to $t_3$ are $2R_3$, $R_1 + R_3$, and $2R_1$, respectively. Therefore, our non-conventional analysis approach can be used for estimating the radial size of MC, propagation speeds of different substructures, and instantaneous expansion speed at a particular instance, even when relying solely on single-point in situ spacecraft measured speeds of different substructures at different instances. In the following, we estimate the propagation speeds, instantaneous expansion speeds, and radial size of the MC using the non-conventional approach at different instances of the arrival of LE, center, and TE at 1 AU.

\subsubsection{Estimates From Non-conventional Approach to Single-point In Situ Observations at Different Instances}\label{sec:insitu_estimate}

\begin{figure*}
    \centering
    \includegraphics[scale=0.6]{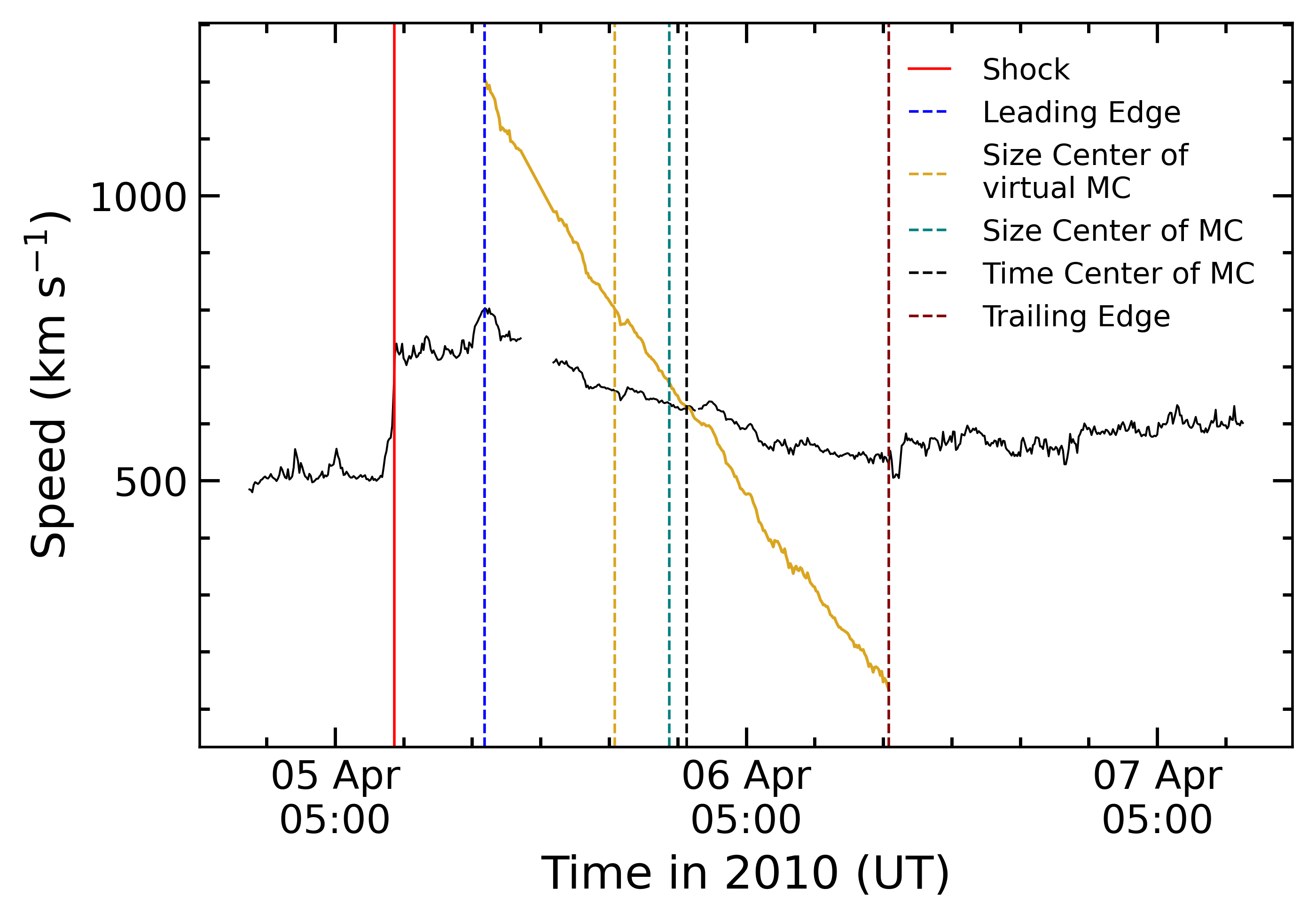}
    \includegraphics[scale=0.6]{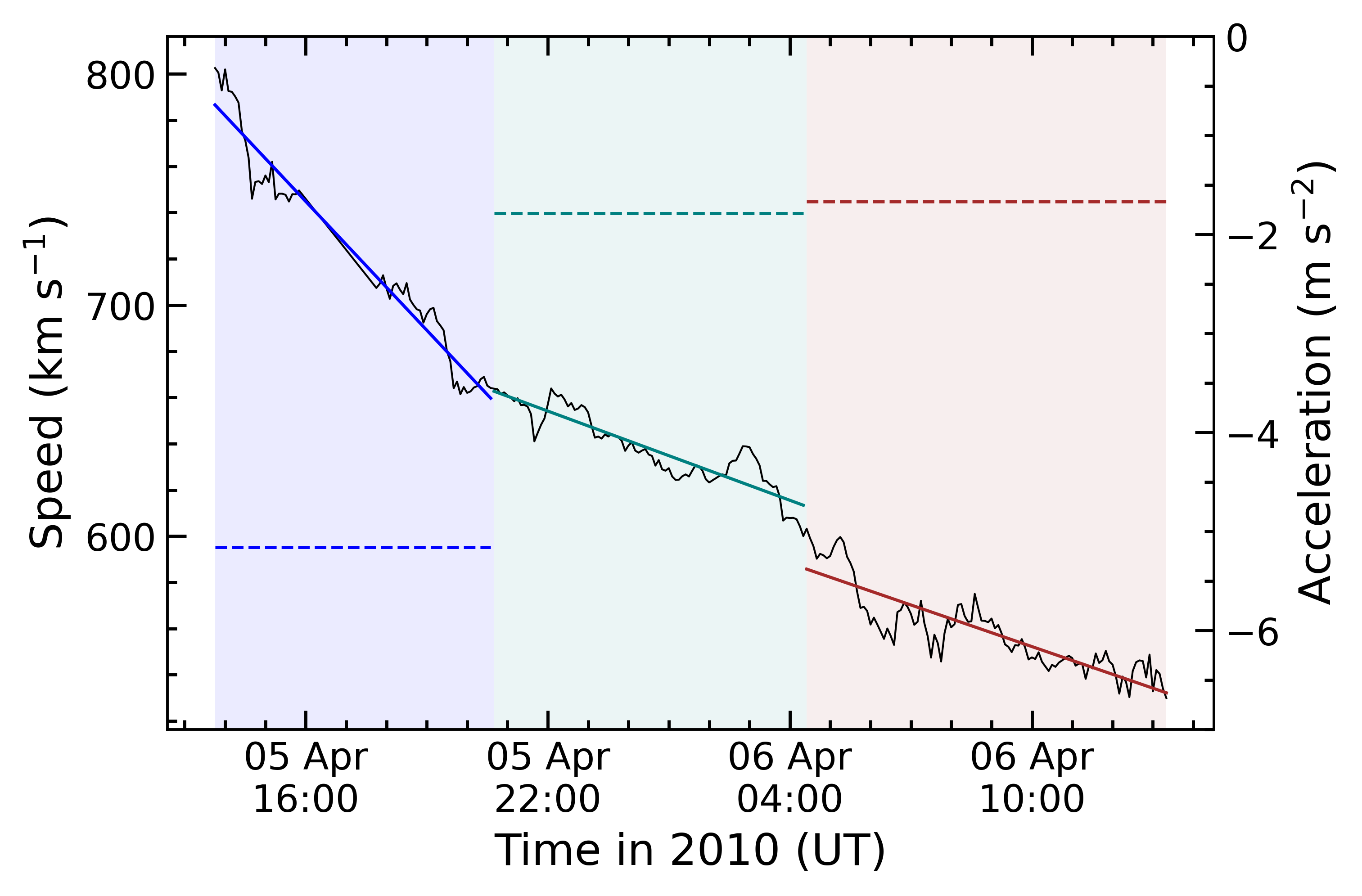}
    \caption{The left panel shows the in situ measured speed profile for the selected CME. The vertical solid red line marks the arrival of the shock associated with the 2010 April 3 CME. The blue, green, black, and maroon vertical dashed lines denote the LE, size center, time center, and TE of the MC, respectively. The yellow solid line denotes the speed profile of the virtual MC with a steeper slope, and the yellow vertical dashed line shows the arrival of the size center of the virtual MC. The right panel shows the in situ measured speed during the MC of the 2010 April 3 CME on the y-axis (left), while the y-axis (right) shows the acceleration. The speed profile is divided into three equal segments (transparent fill areas with blue, green, and maroon) based on the size of the MC. The blue, green, and maroon solid lines represent the linear fitting of the speed profile in each segment for the MC LE, center, and TE, respectively, while the dashed line represents the slope (acceleration) of the linear fitting of the speed.}
    \label{fig:insitu_spe}   
\end{figure*}

 As described in Section~\ref {sec:app}, we refer to the size center as the center. Several earlier studies have considered no difference between the arrival of the time center (arrival time of half of the MC's total duration) and the size center (arrival of half of the MC’s total radial size) of the MC at the in situ spacecraft. This is because such studies assumed the constant expansion speed of the CME during its passage at in situ spacecraft \citep{Owens2005,Gulisano2010,Regnault2024}. We emphasize that the arrival of the size center and time center could often be different, especially for highly accelerating/decelerating CMEs having a larger expansion speed. We note this difference for our selected CME/MC in the left panel of Figure~\ref{fig:insitu_spe}, where the arrival of the size center (green dashed line) at 1 AU precedes the time center (black dashed line) of the MC by approximately 1 hour. The radial size of the MC at the arrival of LE at 1 AU is estimated as 76.8 $R_\odot$ by integrating the speed with time during the MC passage over the in situ spacecraft. This is the most reliable estimation of radial size, considering the nose encounter of the MC with the in situ spacecraft.

We note a difference between the arrival time of the size center and time center for this CME, despite it showing the signatures of minimal deceleration and expansion at the in situ spacecraft. This suggests that the difference between these two centers would be much larger for CMEs experiencing a larger expansion and acceleration/deceleration. Since the conventional method assumes no acceleration while estimating the expansion speeds, their estimates would not be appropriate, especially for such CMEs. Our non-conventional approach would be valid in treating such CMEs because a constant acceleration/deceleration of the CME substructures is already considered for estimating the radial size and speeds at different instances. The selection of this CME was to show that the conventional method is not even reliable for the case where one expects it to be. This CME is one of the candidates to show the best achievable performance of the conventional method in estimating the expansion speed, and the performance of the conventional method for other CMEs with deceleration/acceleration/expansion will worsen further.

We demonstrate a larger difference (4 hours) in the size and time center of a virtual MC as depicted by the yellow solid line in the left panel of Figure~\ref{fig:insitu_spe} profile. The virtual MC has identical boundaries and a time center as the actual MC but has a steeper slope of linearly decreasing speed than the actual MC. A greater disparity between the size and time center of virtual MC suggests its more substantial expansion during MC's total duration \citep{Lugaz2020a}. It becomes apparent the size center and time center for MCs are not synchronous and are dependent on the expansion speed of the MCs. This dependency can be further substantiated through in situ observations of faster-expanding CMEs from spacecraft approaching the Sun.

Moreover, to apply our non-conventional approach to single-point in situ observations, we require the acceleration of LE, size center, and TE on their arrival at 1 AU. However, in the absence of in situ measurements of the same substructures (features) at multiple instances, it is difficult to accurately calculate their acceleration on their arrival at a time. Additionally, \citet{Temmer2022} has shown that LE is not a sharp feature but has some thickness. In this spirit, we consider some thickness of each feature of MC and use the in situ measured speed within the thickness to derive the constant acceleration. We divide the speed profile of the MC into three equal segments based on MC radial size, as shown in the right panel of Figure~\ref{fig:insitu_spe}. The blue, green, and maroon solid lines represent the linear fitting for the speed of the LE, center, and TE segments. The derived constant acceleration for each feature (-5.1 m s$^{-2}$ for LE, -1.8 m s$^{-2}$ for the center, and -1.6 m s$^{-2}$ for TE) is shown with the dashed line of the same color as the corresponding fitting. Using the estimated acceleration of different features in our non-conventional approach, we estimate the propagation speed of the LE, center, and TE at different instances. Furthermore, we estimate the radial size and instantaneous expansion speed of the MC at the arrival of LE (at $t_1$), center (at $t_2$), and TE (at $t_3$) at 1 AU.

The in situ measured speed of the CME LE on its arrival at $t_1$ (13:43 UT on April 5, as shown with a blue vertical dashed line in the left panel of Figure~\ref{fig:insitu_spe}) at 1 AU is 803 km s$^{-1}$. Further, using our non-conventional approach with the first equation of motion (described in Section~\ref{sec:app}), we estimated the speed of different features at different instances where in situ measurements are not available. These estimates are listed in the non-bold font in the last three columns of the middle panel of Table~\ref{tab:tab_1}. From the table, we note that the speed of the CME LE at the arrival of the center and TE at 1 AU using our non-conventional approach is 603 and 365 km s$^{-1}$, respectively. The in situ measured speed of the center on its arrival at $t_2$ (00:30 UT on April 6, as shown with a green vertical dashed line in the left panel of Figure~\ref{fig:insitu_spe}) at 1 AU is 635 km s$^{-1}$ while its estimated speed at the arrival of the LE and TE at 1 AU are 704 and 553 km s$^{-1}$, respectively. The measured speed of the TE on its arrival at $t_3$ (13:20 UT on April 6, as shown with a maroon vertical dashed line in the left panel of Figure~\ref{fig:insitu_spe}) at 1 AU is 530 km s$^{-1}$ while its estimated speed at the arrival of the LE and center at 1 AU are 672 and 607 km s$^{-1}$, respectively. The decrease in speed of LE from $t_1$ to $t_3$ is much larger and becomes even smaller than the speed of the following features (center and TE) at the same instances as shown in the last three columns of the middle panel of Table~\ref{tab:tab_1}. This incorrectly implies the decreasing size of the MC, and therefore, speed estimates for LE are inaccurate. This could be possible because of the overestimated deceleration of LE used in the non-conventional approach. Therefore, the instantaneous expansion speed of the CME at $t_2$ and $t_3$ is estimated using the center and TE propagation speed, as mentioned in the third column of the bottom panel of Table~\ref{tab:tab_1}.

We also estimate the distance traveled by CME features/substructures during any two instances using our non-conventional approach with the second equation of motion as described in Section~\ref{sec:app}. The estimates of distance traveled by the LE from $t_1$ to $t_2$, $t_2$ to $t_3$, and $t_1$ to $t_3$ are $R_2$, 2$R_3$ - $R_2$, and 2$R_3$, respectively, as labeled in Figure~\ref{fig:MC_1Dcut}. The estimated value of $R_2$, 2$R_3$ - $R_2$, and 2$R_3$ are 38.9 $R_\odot$, 31.9 $R_\odot$, and 70.8 $R_\odot$, respectively. The estimates of distance traveled by the center from $t_1$ to $t_2$, $t_2$ to $t_3$, and $t_1$ to $t_3$ are $R_1$ (37 $R_\odot$), $R_3$ (39.2 $R_\odot$), and  $R_1$ + $R_3$ (76.2 $R_\odot$), respectively. The estimates of distance traveled by the TE from $t_1$ to $t_2$, $t_2$ to $t_3$, and $t_1$ to $t_3$ are 2$R_1$ - $R_2$ (35.4 $R_\odot$), $R_2$ (37.5 $R_\odot$), and  2$R_1$ (72.9 $R_\odot$), respectively.

\begin{table}
    \centering
    \footnotesize
    \begin{tabular}{cccc|ccc}
    \hline
     \multicolumn{7}{c}{Arrival Time of CME Features at 1 AU (UT)} \\
    \hline
   \multicolumn{2}{c}{\multirow{2}{6em}{CME Feature}} & \multicolumn{2}{c}{{GCS+SSSE}} & \multicolumn{2}{c}{\multirow{2}{3em}{In Situ}} & {$\Delta$t}\\ 
   \multicolumn{2}{c}{} & \multicolumn{2}{c}{+DBM} & \multicolumn{2}{c}{} & (hr)
   \\
    \hline
   \multicolumn{2}{c}{LE (at $t_1$)} & \multicolumn{2}{c}{5 Apr 14:28} & \multicolumn{2}{c}{5 Apr 13:43} & 0.75 \\
    \multicolumn{2}{c}{Center (at $t_2$)} & \multicolumn{2}{c}{6 Apr 15:22}  & \multicolumn{2}{c}{6 Apr 00:30}  & 14.87 \\
     \multicolumn{2}{c}{TE (at $t_3$)} & \multicolumn{2}{c}{9 Apr 01:37}  & \multicolumn{2}{c}{6 Apr 13:20} & 60.28 \\
    \hline
    \multicolumn{7}{c}{Speed of CME Features at Different Instances (km s$^{-1}$)} \\
    \hline
    \multirow{2}{3em}{} & \multicolumn{3}{c|}{GCS+SSSE+DBM} & \multicolumn{3}{c}{In Situ + Eq. of Motion}  \\
    \hline
    
    Instance & ${V_{LE}}$ & ${V_{center}}$ & ${V_{TE}}$ & ${V_{LE}}$ & ${V_{center}}$ & ${V_{TE}}$ \\
    \cline{2-7}
    \hline
    $t_1$ & 639 & 466 & 293 & \textbf{803} & 704 & 672 \\
    $t_2$ & 596 & 435 & 274 & 603 & \textbf{635} & 607 \\
     $t_3$ & 556 & 406 & 256 & 365 & 553 & \textbf{530} \\
     \hline
    \multicolumn{7}{c}{Expansion Speed at Different Instances (km s$^{-1}$)} \\
    \hline
    \multicolumn{2}{c}{\multirow{2}{3em}{Instance}} & \multicolumn{2}{c}{{GCS+SSSE}} & \multicolumn{2}{c}{{In Situ}} & \multirow{2}{3em}{${\Delta V_{exp}}$}\\ 
    \multicolumn{2}{c}{} & \multicolumn{2}{c}{+DBM} & \multicolumn{2}{c}{+ Eq. of Motion} & \\
    \hline
    \multicolumn{2}{c}{$t_1$} & \multicolumn{2}{c}{173} & \multicolumn{2}{c}{${V_L - V_C}$ = 99} & 74 \\
    \multicolumn{2}{c}{$t_2$} & \multicolumn{2}{c}{161}  & \multicolumn{2}{c}{$V_C - V_T$ = 28}  & 133 \\
    \multicolumn{2}{c}{$t_3$} & \multicolumn{2}{c}{150}  & \multicolumn{2}{c}{$V_C - V_T$ = 23} & 127 \\
     \hline 
    \end{tabular}
    \caption{The top panel shows the arrival time of different features (LE, center, and TE) of 2010 April 3 CME at 1 AU from GCS+SSSE+DBM methods to remote observations, actual arrival time from in situ measurements, and difference in estimates from the remote and in situ observations. The middle panel shows the propagation speed of different features at different instances ($t_1$, $t_2$, and $t_3$) estimated from GCS+SSSE+DBM methods and our non-conventional approach (as described in Section~\ref{sec:insitu_estimate}) to analyzing the in situ measurements or directly from in situ measurements (bold). The bottom panel shows the instantaneous expansion speed of the CME at different instances from GCS+SSSE+DBM methods and our non-conventional analysis approach to in situ measurements, and the difference in estimates from the remote and in situ observations.}
    \label{tab:tab_1}
\end{table}

Moreover, the radial size (2$R_1$) of the MC at the arrival of LE at 1 AU is directly estimated using the in situ measured speed during the passage of the MC (from its LE to TE) at 1 AU. The radial size of the MC is estimated as 76.8 $R_\odot$, which means that the estimate of $R_1$ (MC's radius at $t_1$) is 38.4 $R_\odot$. Furthermore, the estimated values of $R_1$ using our non-conventional analysis approach on the center and TE of the MC are 37 $R_\odot$ and 36.45 $R_\odot$ (2$R_1$ = 72.9 $R_\odot$), respectively (as shown in the last column of Table~\ref{tab:tab_2}). This suggests that $R_1$ using our non-conventional approach on the center and TE of the MC closely matches with the most reliable estimate directly from the in situ measurements of speed and duration of the MC. This shows that our non-conventional analysis approach is reliable if the acceleration of the CME's features/substructures is estimated with reasonable accuracy. Using the estimated or measured acceleration of CME's features, such an approach could provide a more accurate estimate of the radial size and speeds at different instances, especially for CMEs that experience considerable acceleration/deceleration.

\subsection{Analysis Techniques for Remote Observations}\label{sec:rem}

\begin{figure*}
    \centering
    \includegraphics[scale=0.074,trim={25cm 25cm 42cm 69cm},clip]{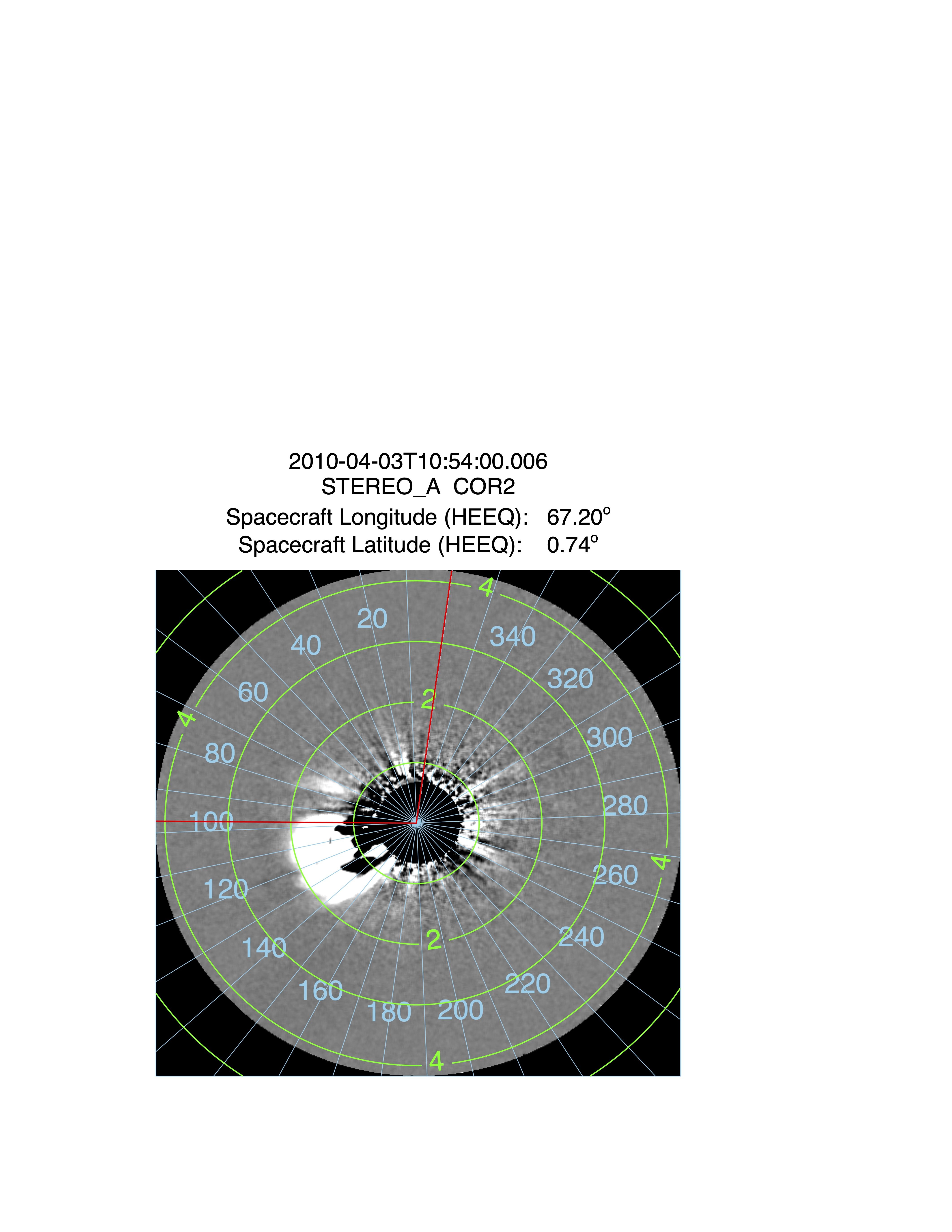}
   \includegraphics[scale=0.074,trim={25cm 25cm 42cm 69cm},clip]{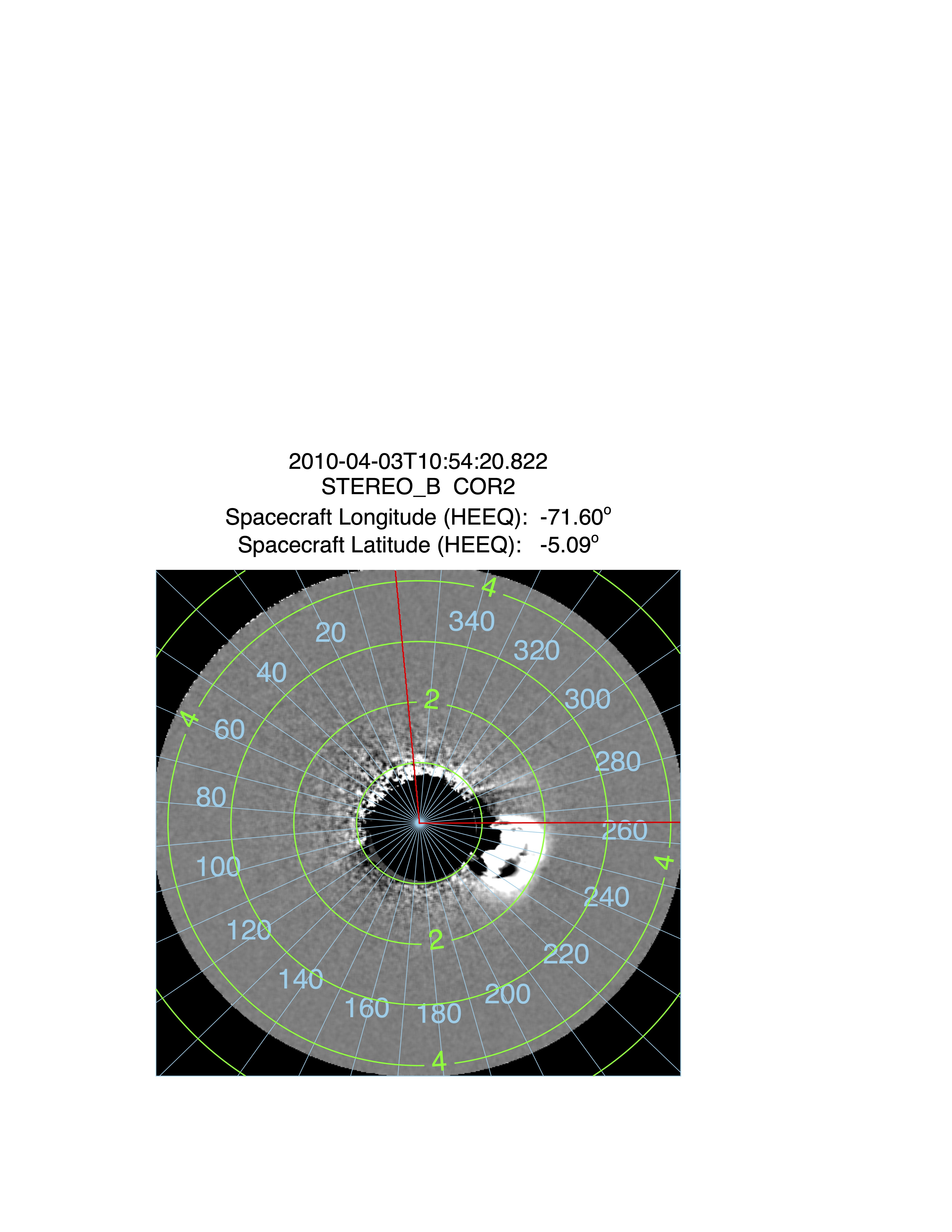}
   \includegraphics[scale=0.074,trim={25cm 25cm 42cm 69cm},clip]{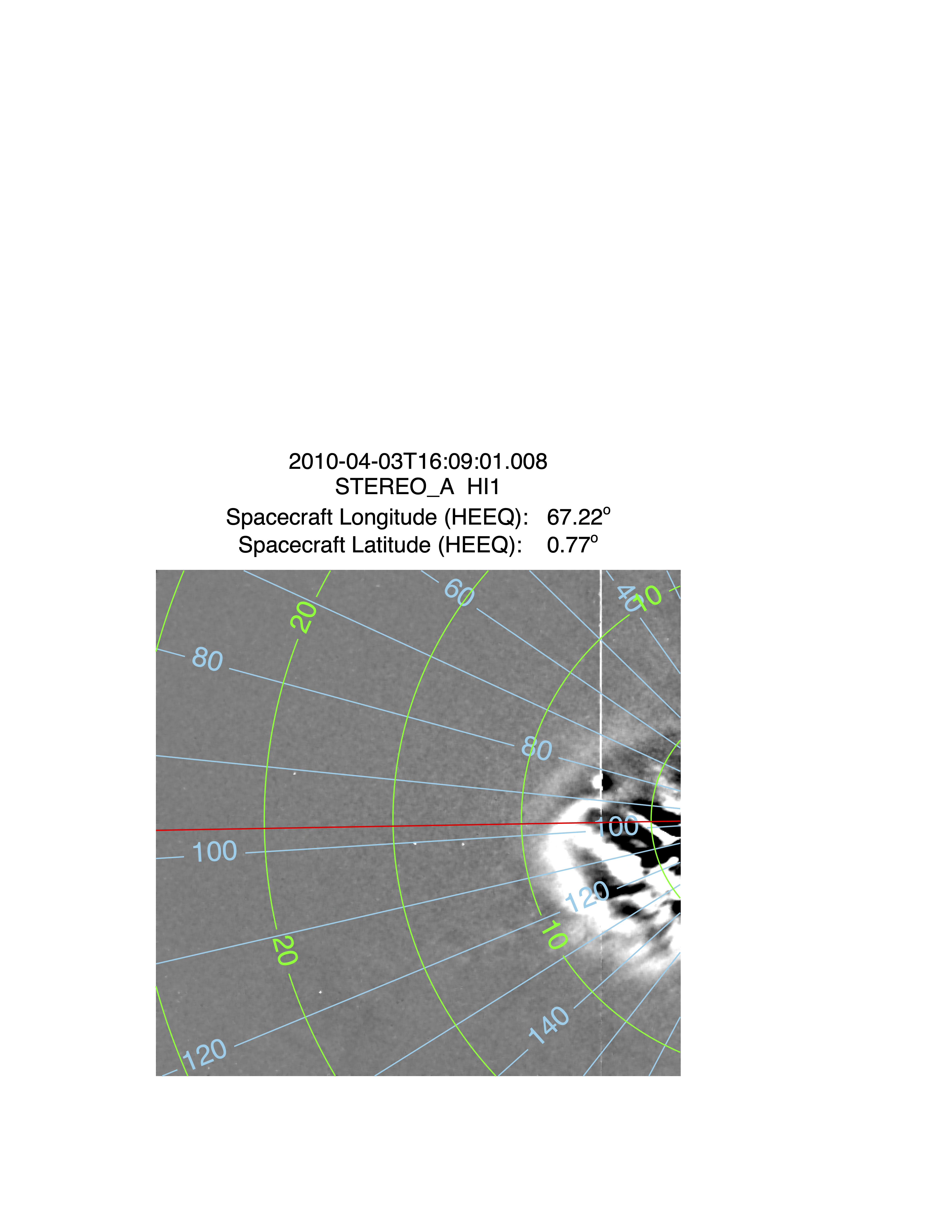}
   \includegraphics[scale=0.074,trim={25cm 25cm 42cm 69cm},clip]{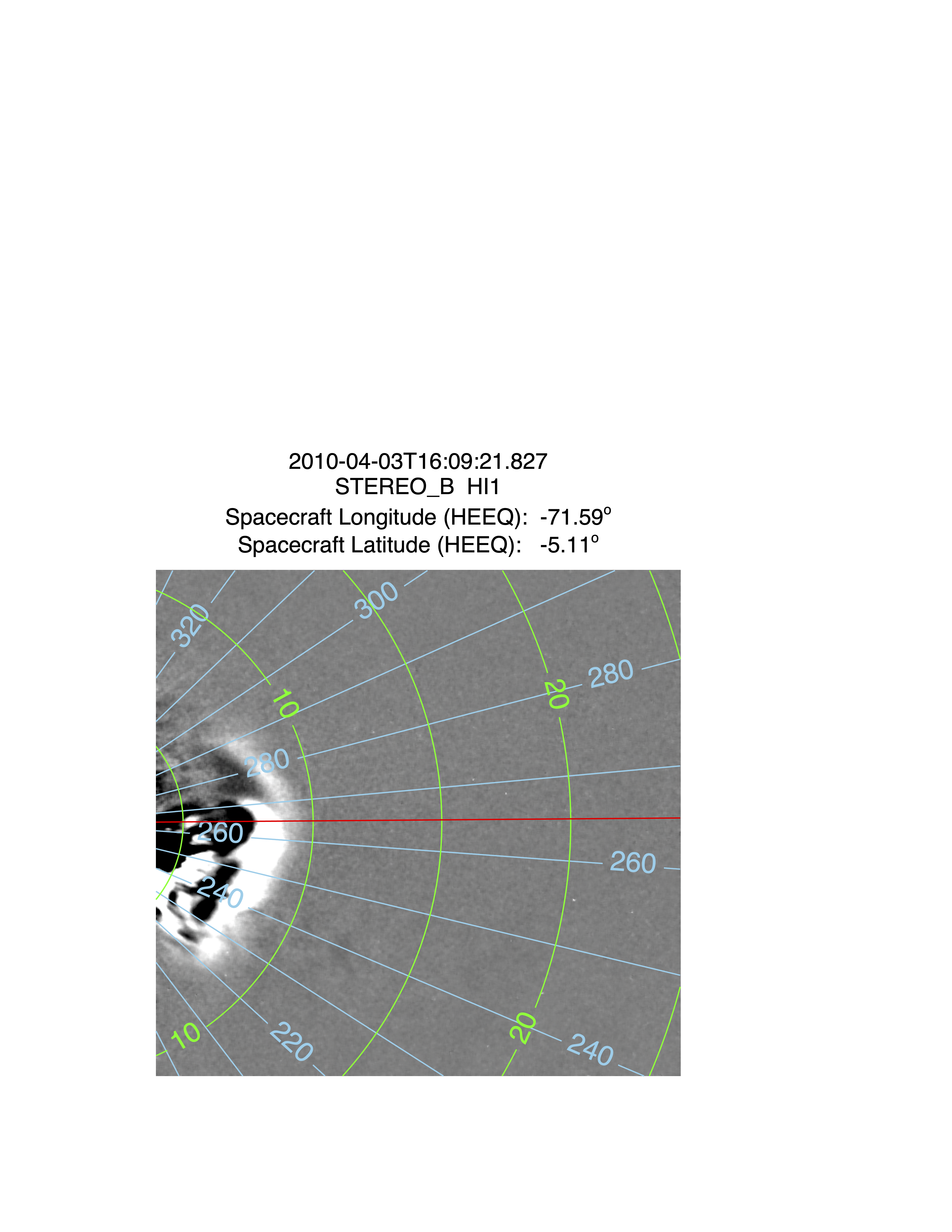}
   \includegraphics[scale=0.074,trim={25cm 25cm 42cm 69cm},clip]{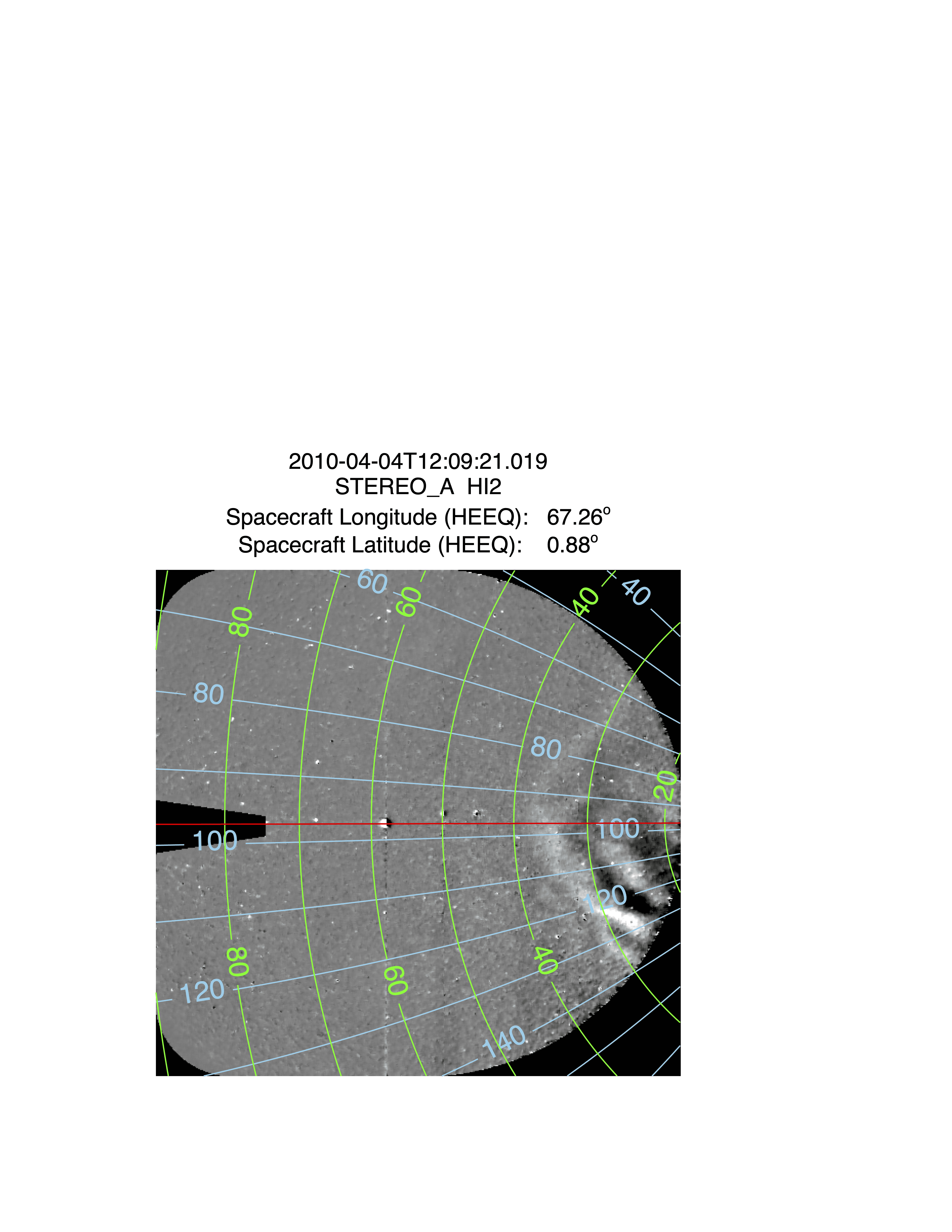}
   \includegraphics[scale=0.074,trim={25cm 25cm 42cm 69cm},clip]{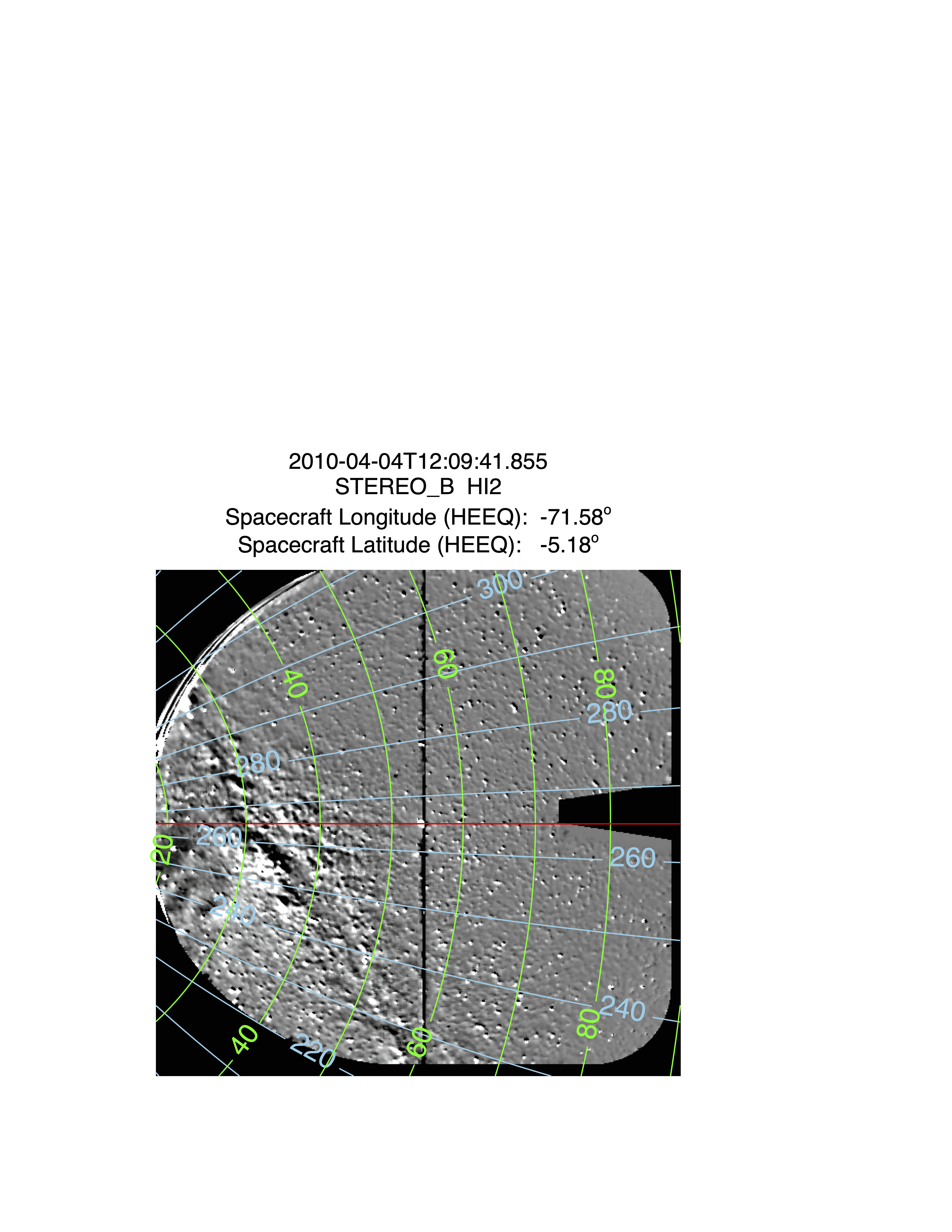}
    \caption{The top to bottom panels show the evolution of 2010 April 3 CME observed in COR2, HI1, and HI2, respectively, from \textit{STEREO-A} in the left column and \textit{STEREO-B} in the right column. The contours of the elongation angle (green) and the position angle (blue) are overlaid on the images. The horizontal red line is at the position angle of the Earth in the ecliptic, while the vertical red line in the top panel marks the zero-degree position angle.}
    \label{fig:CME_ima}
\end{figure*}

The evolution of CME in the COR and HI images is shown in Figure~\ref{fig:CME_ima}. We utilized remote COR and HI observations of the CME and applied conventional reconstruction methods to estimate the CME kinematics. The methods used are Graduated Cylindrical Shell (GCS) model \citep{Thernisien2009} on the COR observations, Stereoscopic Self-Similar Expansion (SSSE) method \citep{Davies2013} on the time-elongation maps (J-maps) constructed from COR and HI observations, and Drag-Based Model (DBM) \citep{Vršnak2013}. The DBM is used only beyond the heights where the CME could not be tracked unambiguously in the HI observations.

\subsubsection{Graduated Cylindrical Shell Model}\label{sec:gcs}

 \begin{figure*}
    \centering
    \includegraphics[scale = 0.44]{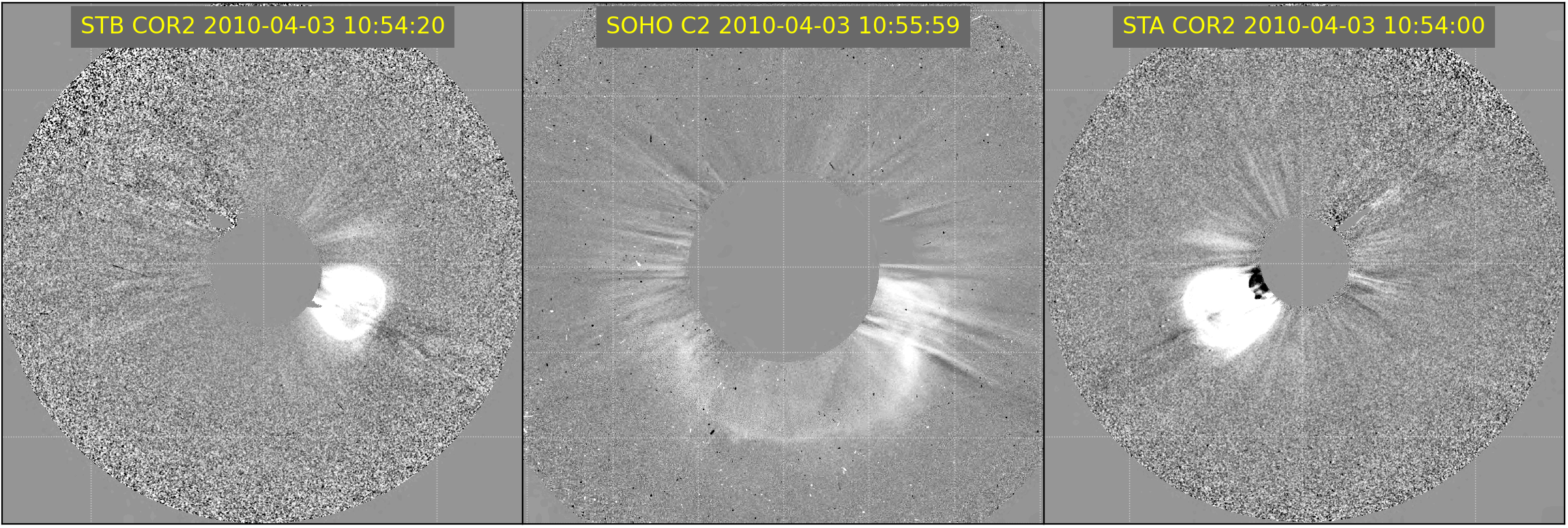}
    \includegraphics[scale = 0.44]{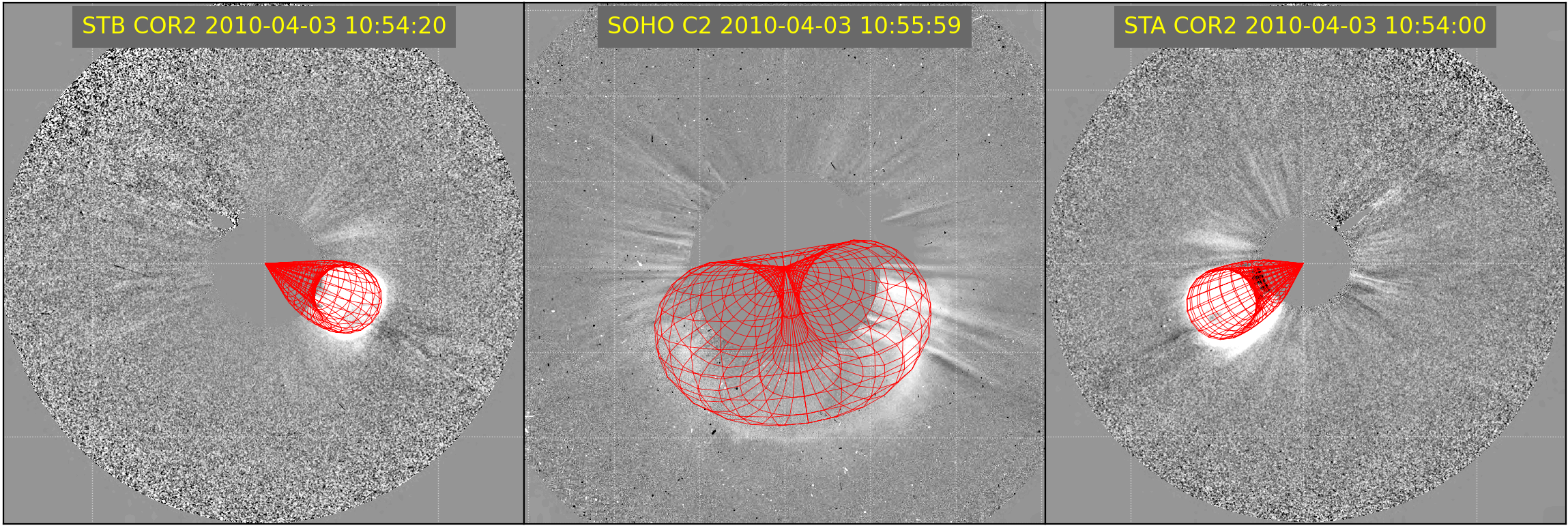}
    \caption{{GCS fitting for 2010 April 3 CME using the simultaneous coronagraphic images from three viewpoints of \textit{STEREO-B}/COR2 (left), \textit{SOHO}/LASCO C2 (center), and \textit{STEREO-A}/COR2 (right). STB and STA represent the STEREO-B and STEREO-A spacecraft, respectively. The top and bottom panels show coronagraphic images without and with GCS model-fitted red wireframes, respectively.}}
    \label{fig:GCS}
 \end{figure*}

 We note that the CME of 2010 April 3 was observed in \textit{SOHO}/LASCO at 10:33 UT and appeared as a partial halo. \textit{STEREO-A}/COR1 and \textit{STEREO-B}/COR1 observed the CME at 09:05 UT in the SE and SW quadrants. For the 3D kinematics of the selected CME, we implement the GCS model \citep{Thernisien2006} on the contemporaneous coronagraphic images from the viewpoints of \textit{SOHO}, \textit{STEREO-A}, and \textit{B}. This model assumes the CME flux rope to be in the shape of a hollow croissant, the structure of which can be adjusted by the six free parameters: latitude ($\theta$) and longitude ($\phi$) of the CME, the half angle between the two conical legs ($\alpha$), tilt angle ($\gamma$), aspect ratio ($\kappa$) and height of the CME leading edge ($h$). The leading edge of this CME is tracked from a height of 2.06 $R_{\odot}$, as observed in the SECCHI/COR1, up to a distance of 13.7 $R_{\odot}$, utilizing the SECCHI/COR2. Figure~\ref{fig:GCS} depicts the GCS fitting in Stonyhurst heliographic coordinate system. At the last tracked height, the other five GCS parameters, $\theta$, $\phi$, $\alpha$, $\gamma$, and $\kappa$ are -24$^\circ$, 3$^\circ$, 25$^\circ$, 9.79$^\circ$ and 0.37, respectively. Our model-fitted morphological and dimensional parameters are consistent with the earlier studies, considering the subjectiveness of the manual fitting using the GCS model \citep{Möstl2010,Liu2011,Mishra2020,Martinic2022}.

\begin{figure*}
    \centering
    \includegraphics[scale=0.80,trim={0.2cm 0.3cm 0.24cm 0.24cm},clip]{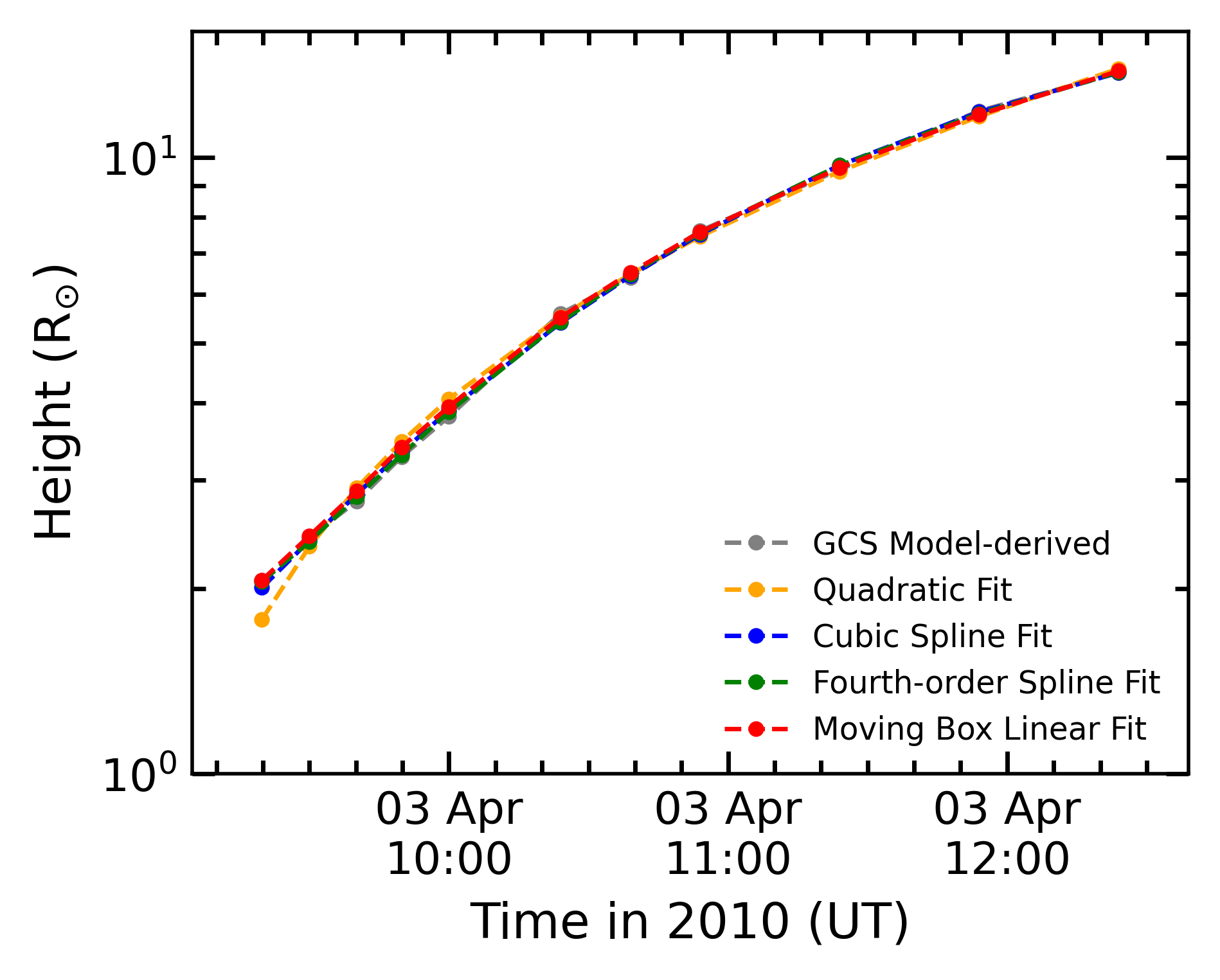}
    \hspace{0.5cm}
    \includegraphics[scale=0.80,trim={0.2cm 0.3cm 0.24cm 0.24cm},clip]{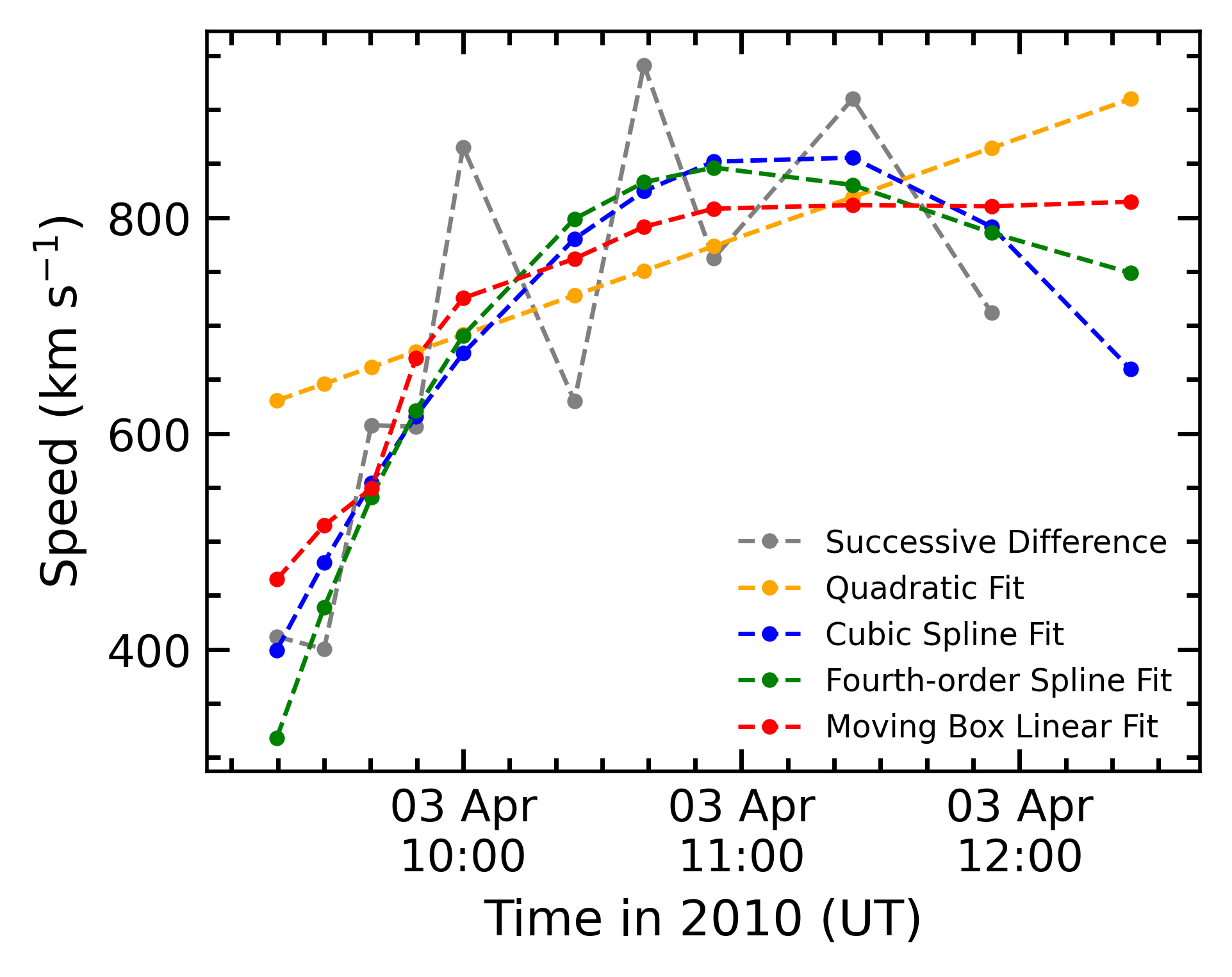} \\
    \vspace{0.3cm}
    \includegraphics[scale=0.032,trim={13.2cm 2.5cm 11.7cm 3.2cm},clip]{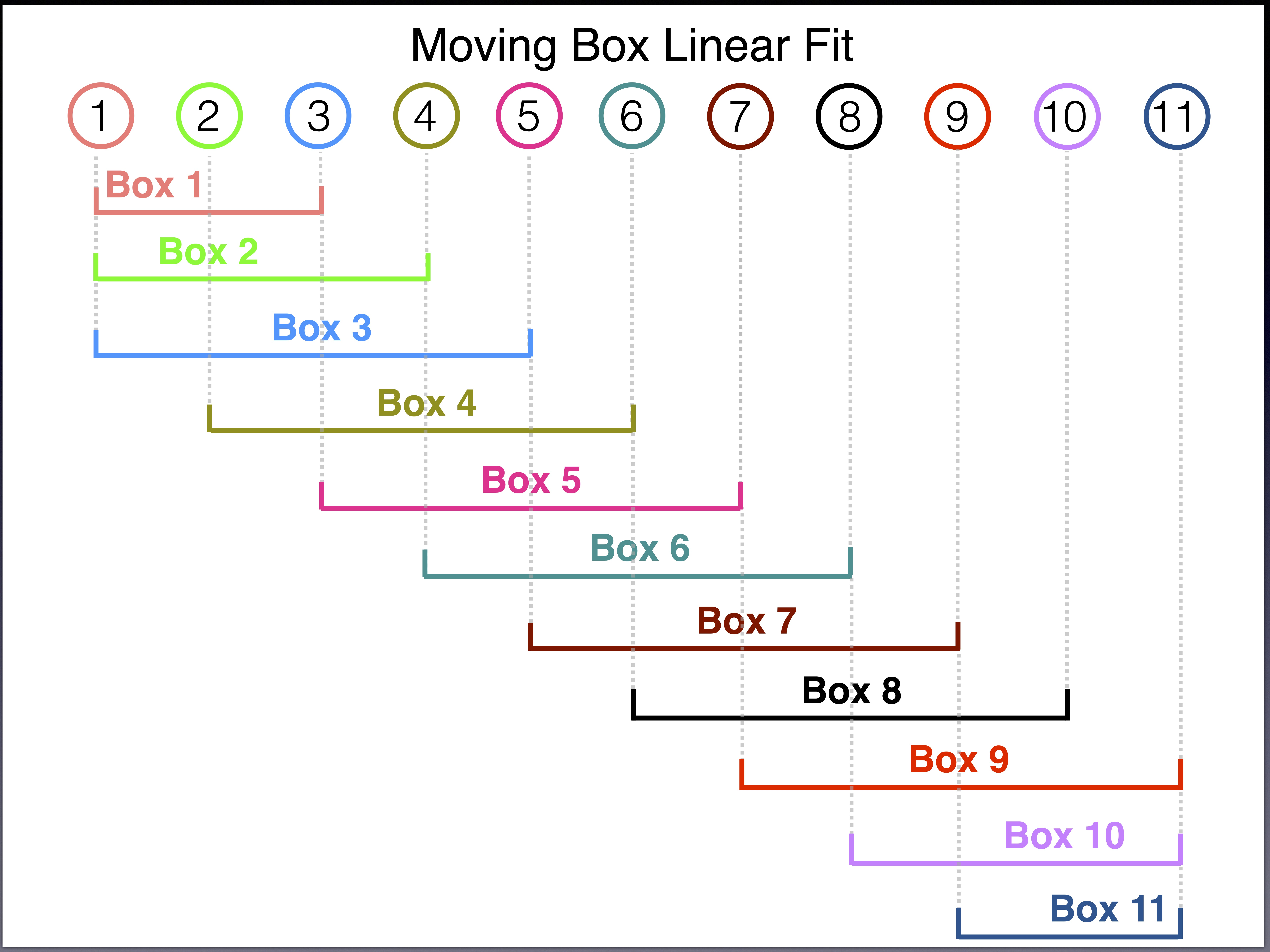} 
    \hspace{0.5cm}
    \includegraphics[scale=0.81,trim={0.2cm 0.3cm 0.24cm 0.1cm},clip]{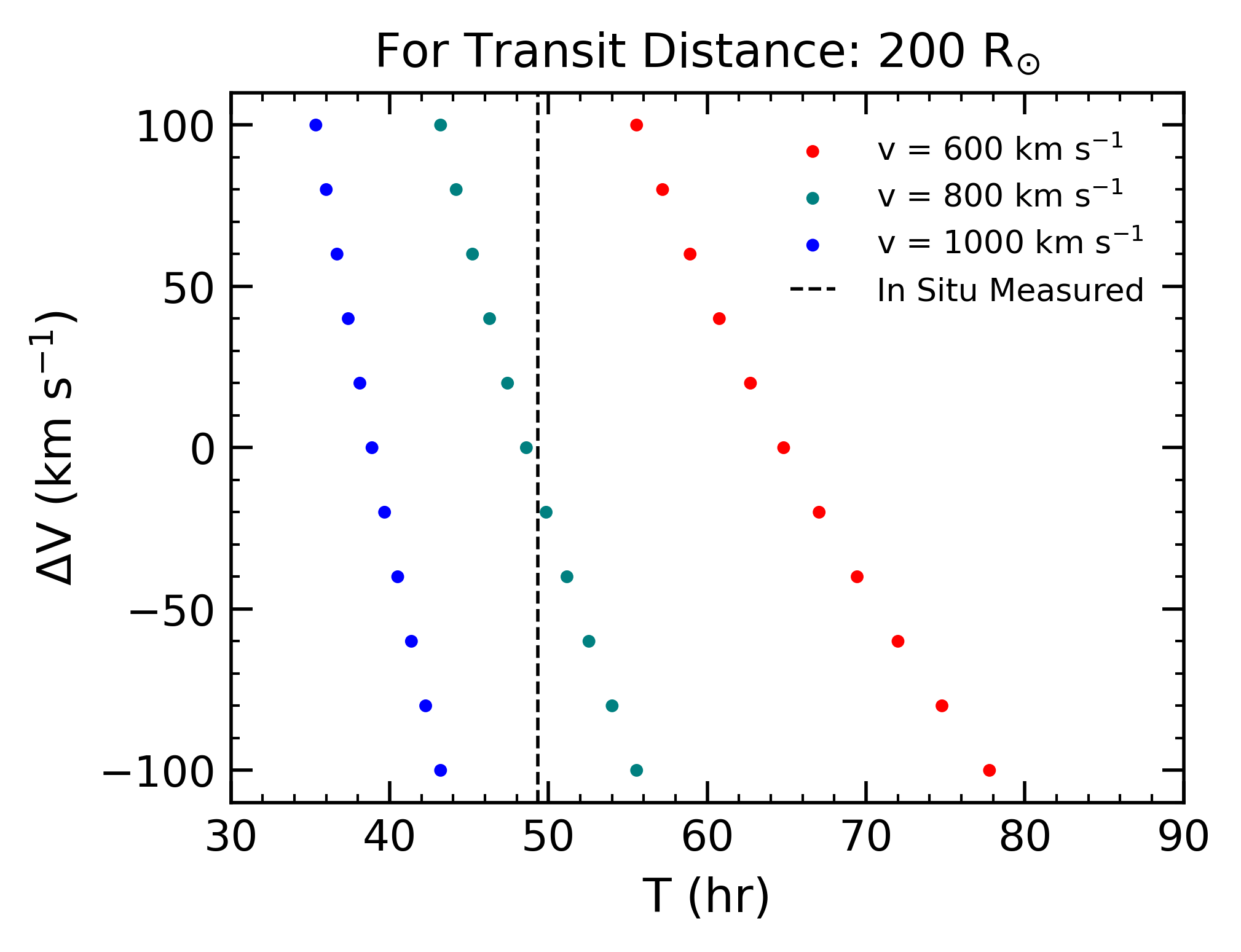}
    \caption{The left panel of the top row shows the height profile from the GCS model-derived in grey. The orange, blue, green, and red show the quadratic fit, cubic spline fit, fourth-order spline fit, and moving box linear fit of GCS model-derived height. The right panel of the top row shows the speed profiles of the respective height-time measurements. The left panel of the bottom row illustrates our adopted moving box linear fit technique. The right panel of the bottom row shows the transit time ($T$) of the CME for the transit distance of 200 $R_\odot$ for different speeds ($\Delta$V). The red, green, and blue correspond to three cases of speeds as 600, 800, and 1000 km s$^{-1}$, respectively, with the uncertainties of $\pm$100 km s$^{-1}$. The vertical black dashed line marks the in situ measured transit time of the CME.}
    \label{fig:diff_tt}  
\end{figure*}

It is always tricky to estimate the evolving speed of a CME from the discrete (largely spaced) measurements of its height and time profile \citep{Liu2010,Mishra2013,Colaninno2013}. The shape of the speed (i.e., derivative) profile can often vary by adopting different functions to fit the measured height-time data points. We examine the effects of using different methods to estimate the speed from the GCS model-derived height-time evolution shown in the grey in the left panel of the top row of Figure~\ref{fig:diff_tt}. This plot also shows the different fitting to the GCS model-derived height measurements, such as quadratic fitting (orange), cubic spline fitting (blue), and fourth-order spline fitting (green). The right panel of the top row shows the different speed profiles from various fittings. The magnitude and trend of CME speeds from the different fitting techniques are significantly different. These speeds are the CME LE speed. We note that the polynomial fitting used for the entire duration of the CME evolution can remove the actual short-term fluctuations from the speed profile. In contrast to polynomial fit, making a successive difference (shown in grey) of height-time points for the derivative can bring unphysical fluctuations in the speed of a CME. This implies the uncertainties involved in the kinematics despite the accuracy of CME 3D height measurements from the GCS model.

As a compromise, to retain a possible real change in the CME speed over a few hours, we also use a moving box linear fitting to the smoothed height-time points (shown in red in the left panel of the top row of Figure~\ref{fig:diff_tt}) and derive the speed from the slope of the linear fit. The number of data points (window size) used for the smoothing and linear fit for estimating the speed at a particular instance is shown by the schematic in the left panel of the bottom row of Figure~\ref{fig:diff_tt}. In this panel, circles with different colors, marked with numbering, show the sequence of measured heights corresponding to different times. The panel also shows the sequence of moving boxes, with different colors, consisting of several data points over which smoothing and linear fit are done. The speed at any data point marked with a certain color/number is estimated using the smoothing box of the same color/number. The number of data points in each moving box is shown, which in our case is smaller at both ends to get the number of speed points equal to the number of data points in the heights. The speed derived from the moving box linear fit (red) is also shown in the right panel of top row of Figure~\ref{fig:diff_tt}.

We notice different speed profiles, especially at both ends of the data points, in the right panel of the top row of Figure~\ref{fig:diff_tt} from different fittings of height-time measurements. In the beginning ($\sim$ 2 $R_\odot$), the speed is in the range of around 300 to 650 km s$^{-1}$ while it is in the range of around 650 to 950 km s$^{-1}$  at the last tracked height ($\sim$13.7 $R_\odot$). Such a difference in the speeds shows the possible uncertainties involved in estimating the transit time of the CME for the transit distance (distance between the last tracked height of the CME LE and the L1 point) of around 200 $R_\odot$ from the coronagraphic height to L1 point. The right panel of the bottom row of Figure~\ref{fig:diff_tt} shows transit time ($T$) on the X-axis, taken by the CME for a transit distance of 200 $R_\odot$ at various speeds shown on the Y-axis. The transit time is determined by employing a constant speed at the last tracked height to encompass the specified distance (200 $R_\odot$). It is clear that uncertainties in the speed by $\pm$100 km s$^{-1}$ can give rise to 10-20 hours of error in arrival time for a typical fast and slow speed CME. This plot depicts that the change in speed by equal magnitude ($\pm$100 km s$^{-1}$) gives the non-equal change in the arrival time. Therefore, it suggests that the change in arrival time due to a change ($\pm$100 km s$^{-1}$) in speed is smaller for faster-speed CMEs. This CME is well identified in the in situ observations at 1 AU, and its in situ measured arrival time (arrival time of the CME LE identified from in situ magnetic field and plasma measurements) is at 13:43 UT on April 5. Based on the in situ measurements, the CME transit time is around 49.32 hours, marked with a vertical black dashed line, implying an average CME speed of around 800 km s$^{-1}$ between the last tracked height and 1 AU. It should be noted that comparing arrival times from remote and in situ measurements can bring additional inconsistency because the CME radial propagation direction estimated from remote observations is not strictly in the ecliptic plane where in situ measurements are taken. It could be possible to correct the 3D speeds to get its component along the Sun-spacecraft line in the ecliptic plane before comparing with in situ measurements, but such a correction is not done for the selected CME as it will not much affect the estimates because the propagation of the selected CME is closely along the Sun-Earth line.

We also estimated the radius of the CME flux rope as $R = \left(\frac{\kappa}{1 + \kappa}\right)h$ where $\kappa$ is the aspect ratio derived from the GCS model and $h$ is the 3D height of the LE. The continuous evolution of radial expansion speed is estimated using the radius of the flux rope. The height of the size center and TE of the CME are estimated to be $h - R$ and $h - 2R$, respectively. Using estimated size center and TE heights, we obtain the size center and TE speed by employing the moving box linear fit technique. The 3D kinematics plot for the CME's different features (LE, center, and radius) is shown in Figure~\ref{fig:kin}. This figure, from the top to bottom panels, shows the height, speed, and acceleration of different features. The unfilled circle is used to denote the estimates from the GCS model, and different colors (blue, green, and yellow) are used to mark the different features of the CME (LE, size center, and radius). The error in the derived kinematics of each feature is represented by transparent fill areas over the data points with the same color as used for the data points of the corresponding feature. The error bars are derived by considering an error of 10$\%$ in the measurements of the height at each data point.

From the uncertainties in the derived speeds, it is clear that the 3D speed of the CME at coronagraphic heights and assuming it to be constant for the remaining interplanetary journey of the CME can bring large errors in estimating the arrival time of even the CME LE. This can also bring errors in the estimates of expansion speed and radial sizes of the CME. Therefore, to make a reasonable comparison of the estimates from remote observations with those from in situ observations, we further examine the continuous evolution of different substructures of the CME and derive its radial size at varying distances from the Sun. For tracking the CME beyond \textit{STEREO}/COR2 field of view (FOV), we utilize the HI1 and HI2 observations and implement the SSSE method to derive the 3D kinematics that further has been used as inputs to DBM, as described in Section~\ref{sec:ssse}.

\subsubsection{Implementing SSSE Method on \textit{STEREO}/HI Observations} \label{sec:ssse}

The tracking of LE of a CME in HI observations has often been done using J-maps, i.e., time-elongation maps \citep{Sheeley1999,Davies2009,Mishra2013}. We constructed the J-maps along the ecliptic plane using the running difference images from the COR2, HI1, and HI2 onboard \textit{STEREO-A} and \textit{B}. The details of the procedure to construct the J-maps are exactly the same as described in \citet{Mishra2013}. We tracked the CME LE using J-maps and derived its elongation-time profile. We employed the SSSE reconstruction method developed by \citet{Davies2013} to estimate the 3D height-time profile of the CME LE. SSSE method treats the CME cross section as a circle in the ecliptic plane with a certain half-angular width ($\lambda$), which propagates self-similarly away from the Sun. In our study, the half-angular width of the CME is derived from the GCS parameters, represented as $\lambda = \alpha \cos(\gamma) + \delta$, where $\delta = \sin^{-1}\kappa$ signifies the thickness of the legs of the CME's hollow croissant shape. The tilt angle of the CME affects the angular width solely through the $cos(\gamma)$ factor, influencing the angle between the legs ($\alpha$). The changes in tilt do not impact the thickness of the legs in the ecliptic plane. An earlier study has shown that the assumed angular width and geometry of the CME, taken as inputs to the model deriving kinematics, can have a significant effect on the estimated kinematics and arrival time of the CME \citep{Rollett2016}. Based on GCS model estimates, in our study, the calculated $\lambda$ approximates 40$^{\circ}$ for the CME of 2010 April 3.

\begin{figure*}
    \centering
    \includegraphics[scale = 0.45]{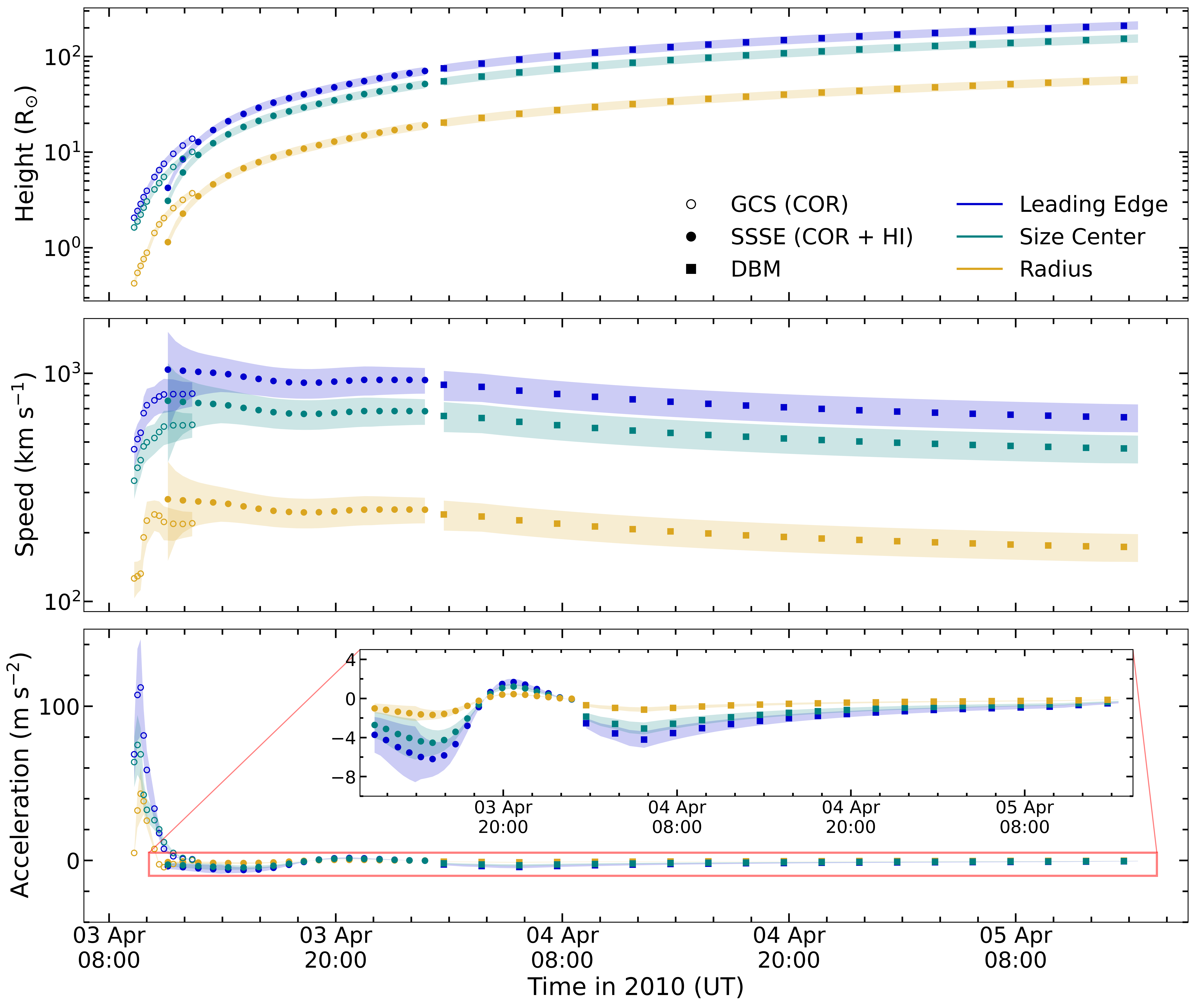}
    \caption{The top to bottom panels show the 3D height, speed, and acceleration of the 2010 April 3 CME LE, size center, and radius up to the instance of CME LE arrival at 1 AU. In each panel, the LE, size center, and radius are shown in blue, green, and yellow colors. The unfilled circles, filled circles, and filled squares represent the estimates from the 3D reconstruction methods GCS, SSSE, and DBM, respectively. The height and speed panel are shown in the log scale. The inset plot in the acceleration panel shows the acceleration of the CME LE, size center, and radius estimated using SSSE and DBM. The error bar over each data point is shown with transparent fill areas of the same color as the corresponding data point.}
    \label{fig:kin}
\end{figure*}

 We applied the SSSE method and derived the 3D height of the CME from 4.2 $R_\odot$ to 70.8 $R_\odot$. The limited tracking of the CME is because it becomes unidentifiable in the HI2 images of \textit{STEREO-B} due to interference from the intense background Milky Way galaxy. For estimating the 3D speed and acceleration from height-time measurements, we use the moving box linear fitting technique described in Section~\ref{sec:gcs}. The obtained 3D kinematics is shown in Figure~\ref{fig:kin}. The speed of the CME LE at 70.8 $R_\odot$ is around 935 km s$^{-1}$. We used the CME LE height from the SSSE method and estimated the radius of the CME as described in Section~\ref{sec:gcs}. Since the aspect ratio value was noted to be constant ($\kappa$ = 0.37) at the last four points derived from the GCS model on coronagraphic images, we assume it to be constant in the HI FOV. CMEs achieving a constant aspect ratio in the low corona (within 10 $R_\odot$) have also been reported earlier \citep{Cremades2020}. The estimated heights of CME LE, size center, and radius of the CME are shown in the top panel of the figure. The filled circle represents the estimates from the SSSE method, and different colors (blue, green, and yellow) mark the different features of the CME (LE, size center, and radius). The speeds and acceleration of these features are shown in the middle and bottom panels, respectively. Our findings confirm the earlier studies that CME shows minimal deceleration in the interplanetary (IP) medium, and it is suitable for our study of the continuous evolution of CME's radial size \citep{Möstl2010,Liu2011,Mishra2014}. We further estimate the evolution of CME beyond the tracked height in the J-maps to 1 AU using the DBM \citep{Vršnak2013} as explained in the following section.

\subsubsection{Implementing DBM for Tracking the CME up to 1 AU}

The SSSE method provided the 3D kinematics of CME up to a distance of around 70.8 $R_\odot$, and we assume that beyond this height, the speed of the CME can be governed primarily by drag forces. Consequently, the acceleration of the CME can be described as drag acceleration, denoted as $a= -\gamma(v-w)|v-w|$, where $v$ represents the CME speed, $w$ is the solar wind speed, and $\gamma$ denotes the drag parameter. The drag parameter is expressed as $\gamma = \frac{c_d A \rho_w}{M}$, incorporating the dimensionless drag coefficient ($c_d =1$), the cross-sectional area of the CME perpendicular to its direction of propagation ($A$), the mass density of the ambient solar wind ($\rho_w$), and the mass of the CME ($M$). We estimate the 2010 April 3 CME drag parameter by estimating its de-projected mass, cross-sectional area, and ambient solar wind density.

Using the theory of Thomson scattering \citep{Billings1966, Vourlidas2000}, the true (de-projected) mass of the CME at the outer edge of COR2 is calculated as $7.98 \times 10^{15} g$. The estimated mass of the CME in our study is consistent (within 25\%) with that estimated in \citet{Temmer2021}. The mass density of the ambient solar wind at various heights ($h$) is computed using the solar wind density model of \citet{Leblanc1998}. We determine the cross-sectional area as $A$ = $\pi(\lambda h)^2$ of the CME employing a half angular width ($\lambda$ = $40^\circ$) derived from the GCS model. By incorporating all the relevant values for drag parameter ($\gamma$) estimation, we ascertain its value to be $0.36 \times 10^{-7}$ km$^{-1}$ for the CME observed on 2010 April 3.

For the specific 2010 April 3 CME under consideration, the input parameters for DBM are the take-off speed of CME $v_0 = 935$ km s$^{-1}$ at $h_o = 70.8$ $R_\odot$, and ambient solar wind speed as $w = 500$ km s$^{-1}$. Although it is difficult to estimate the realistic value of background solar wind speed into which CME has traveled, we, for simplicity, take its average value in the time window of approximately 2 hours before the in situ measured arrival of the CME shock at 1 AU. Our choice of ambient solar wind speed is consistent with the empirically obtained value in the statistical study of \citet{Vršnak2013}, and a similar approach is also taken in \citet{Mishra2013}. The resulting height-time evolution for the LE, size center, as well as the radius of the CME up to 1 AU, is shown in the top panel of Figure~\ref{fig:kin}. We consider that the aspect ratio of the CME remains the same as taken during the evolution in HI1. The speed and acceleration from these estimated heights are derived by applying the moving box linear fit technique described in Section~\ref{sec:gcs} and are shown in the middle and bottom panels. The filled square represents the estimates derived from the DBM method, while distinct colors (blue, green, and yellow) denote different features of the CME (LE, size center, and radius). The transparent fill areas over the data points in corresponding colors indicate the error in the derived kinematics for each feature.

Figure~\ref{fig:kin} shows the 3D kinematics of the CME LE (blue) and center (green) as well as the time evolution of the CME radius (yellow) continuously from the beginning until the CME LE arrives at 1 AU. It is noted that while CME LE arrives at 1 AU (at 14:28 UT on 05 April), the center and TE are at 155.5 $R_\odot$ and 98 $R_\odot$, respectively. To calculate the arrival time of CME's center and TE at 1 AU, we extended the DBM run for heights of CME LE beyond 1 AU. We find that while the center arrives at 1 AU (at 15:22 UT on 06 April), the CME LE and TE are at 291.8 $R_\odot$ and 134.2 $R_\odot$, respectively. On the arrival of CME TE at 1 AU (at 01:37 UT on 09 April), the CME LE and center are at 463.2 $R_\odot$ and 338.1 $R_\odot$, respectively. It is evident that CME's different substructures (LE, center, and TE) are well separated and arrive at different instances ($t_1$, $t_2$, and $t_3$) at 1 AU as estimated from remote observations (GCS+SSSE+DBM). The arrival times estimates from GCS+SSSE+DBM are listed in the second column of the top panel of Table~\ref{tab:tab_1}, which are different than those obtained directly from in situ observations. At the distinct arrival times ($t_1$, $t_2$, and $t_3$) of each feature at 1 AU, the estimated propagation speeds of the CME LE, center, and TE are shown in the second, third, and fourth column of the middle panel of Table~\ref{tab:tab_1}. The expansion speed of the CME at different instances is listed in the second column of the bottom panel of Table~\ref{tab:tab_1}. The distance traveled by any feature (LE, center, and TE) between two instances ($t_1$ to $t_2$, $t_2$ to $t_3$, and $t_1$ to $t_3$) is listed in the second column of Table~\ref{tab:tab_2}. In the following, we will describe the disparity in the estimates from remote and in situ observations.

\subsection{Comparison Between Estimates from Remote and Non-Conventional Approach to In Situ Observations} \label{sec:com}

In this section, the arrival time of different features at 1 AU, their propagation and expansion speeds, and the radial size of the selected MC derived from remote observations combined with DBM are compared with those from the non-conventional approach to the single-point in situ observations.

\subsubsection{Arrival Time of Different Features of the MC at 1 AU} \label{sec:arrtim}

The tracking and estimation of 3D kinematics (from GCS+SSSE+DBM) of different features/substructures of the CME/MC are described in Section~\ref{sec:rem}. We compare the estimated arrival time of the MC features (LE, center, and TE) using remote observations with in situ observations. The top panel of Table~\ref{tab:tab_1} lists the different features of the MC, the arrival time of each feature from the combination of GCS+SSSE+DBM, the arrival time of each feature measured in situ at 1 AU, and the difference between both arrival time ($\Delta t = t_{remote} - t_{in~situ}$). We note that the arrival of LE, center, and TE (i.e., all the features) from 3D kinematics is later than the in situ measured arrival time. The difference in arrival time $\Delta t$ from remote and in situ is 0.75, 14.87, and 60.28 hours for LE, center, and TE, respectively. This clearly shows the challenges involved in accurately estimating the arrival time of the center and TE of the MC, even if the arrival of its LE is reasonably well estimated. The increasingly larger value of $\Delta t$ for the following features of the MC could be due to an underestimation of their propagation speeds from GCS+SSSE+DBM and, consequently, the overestimation of their expansion speeds. The discussion on the estimates of speeds is as follows.

\subsubsection{Propagation and Expansion Speeds of Different Features of the MC at Different Instances}

From the middle panel of Table~\ref{tab:tab_1}, we note the speeds of different features/substructures of the MC at different instances derived from the remote (GCS+SSSE+DBM) and in situ. The in situ measured speeds of LE (at $t_1$), center (at $t_2$), and TE (at $t_3$) on their arrival to 1 AU are in bold font. We notice a large inconsistency between the speeds of the features (especially TE) from the 3D kinematics (GCS+SSSE+DBM) and measured (bold font) in situ. The propagation speeds derived from remote observations seem to be underestimated than those measured in situ. The speeds of features at instances when they are not at 1 AU are estimated using our non-conventional approach with the first equation of motion to in situ observations (as described in Section~\ref{sec:app}), and they are shown in the normal font in the last three columns of the middle panel of the table.

The bottom panel of Table~\ref{tab:tab_1} shows the instantaneous expansion speed at different instances ($t_1$, $t_2$, and $t_3$). The instantaneous expansion speed is calculated as the difference in the propagation speeds of two adjacent features. Since the estimated in situ speed of LE from the non-conventional approach has large errors, it is not used for the calculation of instantaneous expansion speed at $t_2$ and $t_3$. The expansion speed estimated from both remote observations (GCS+SSSE+DBM) and in situ measurements (non-conventional) shows the expected trend of decrease in the expansion speed from $t_1$ to $t_3$. The difference in instantaneous expansion speed from remote and in situ could be possible due to uncertainties in estimates from kinematics and/or in situ. The uncertainties in the estimates can arise due to acceleration derived from in situ measurements and taken as input to the first equation of motion, utilization of DBM taking inputs from the SSSE, and assumption of a constant aspect ratio throughout the interplanetary journey of CME.

\vspace*{-1mm}

\subsubsection{Radial Size of the MC at Different Instances and the Distance Traveled by Different Features of the MC During Two Instances}

The radial size of the MC at different instances and the distance traveled by each feature (LE, center, and TE) of the MC during two instances (as illustrated in Figure~\ref{fig:MC_1Dcut}) using our non-conventional approach to single-point in situ measurements are described in Section~\ref{sec:app} and \ref{sec:insitu_estimate}. Table~\ref{tab:tab_2} shows the radial sizes and distances traveled as estimated from remote (GCS+SSSE+DBM) and in situ (non-conventional approach) observations. From this table, we note a large inconsistency between the findings from remote and in situ observations, especially for features traveled between $t_2$ to $t_3$ and $t_1$ to $t_3$.

\begin{table}
    \centering
    \begin{tabular}{lccc}
    \hline
    \multirow{3}{8em}{CME Feature During Any Two Instances } & \multicolumn{3}{c}{Distance traveled ($R_\odot$)}  \\
    \cline{2-4}
     & \multirow{2}{3em}{Derived} & GCS+SSSE & In Situ \\
     &  & +DBM & +Eq. of Motion \\ 
    \hline
   LE: $t_1$ to $t_2$  & $R_2$ & 78.7 & 38.9 \\
    LE: $t_2$ to $t_3$  & 2$R_3$ - $R_2$ & 171.4 & 31.9 \\
    LE: $t_1$ to $t_3$  & 2$R_3$ & 250.1 & 70.8 \\
    Center: $t_1$ to $t_2$  & $R_1$ & 57.5 & 37 \\
    Center: $t_2$ to $t_3$  & $R_3$ & 125.1 & 39.2 \\
    Center: $t_1$ to $t_3$  & $R_1$ + $R_3$ & 182.6 & 76.2 \\
    TE: $t_1$ to $t_2$  & 2$R_1$ - $R_2$ & 36.2 & 35.4 \\
    TE: $t_2$ to $t_3$  & $R_2$ & 78.8 & 37.5 \\
    TE: $t_1$ to $t_3$  & 2$R_1$  & 115 & 72.9 \\
    
    \hline
    \end{tabular}
    \caption{The table lists the distance traveled by different features (LE, center, and TE) of the CME during any two instances at the passage of the CME features at 1 AU. The estimates of the traveled distance are listed as mathematically derived from GCS+SSSE+DBM methods to remote observations and from our non-conventional approach to analyzing in situ measurements.}
    \label{tab:tab_2}
\end{table}

 The mismatch between the findings from remote (GCS+SSSE+DBM) and those from in situ observations could be possible for several reasons: (i) the uncertainty in the estimated 3D kinematics due to ideal assumptions in the models (GCS+SSSE+DBM), (ii) the assumption of constancy of aspect ratio of the MC during its continuous evolution \citep{Savani2011}, and (iii) the uncertainty in the acceleration calculated from in situ measurements and assuming its constancy in our non-conventional analysis. Since the arrival time estimated using 3D kinematics of the LE closely matches the in situ measurements (Section~\ref{sec:arrtim}), the uncertainties from the model's (GCS+SSSE+DBM) assumptions are expected to be minimal. Also, examining the effect of acceleration on the radial size would require multi-point in situ observations of the same feature at two instances \citep{Lugaz2020a,Regnault2024}, which is not available for the selected CME. It would be interesting to compare the aspect ratio calculated from the single-point in situ observations of our selected MC and compare that from the remote observations. The constancy of the aspect ratio leading to a large disparity in the estimated radial size of the CME is described below.

\subsection{Implications of Evolution of Aspect Ratio of the CME on its Radial Size}\label{sec:aspect}

\begin{figure}
    \centering
    \includegraphics[scale = 0.45]{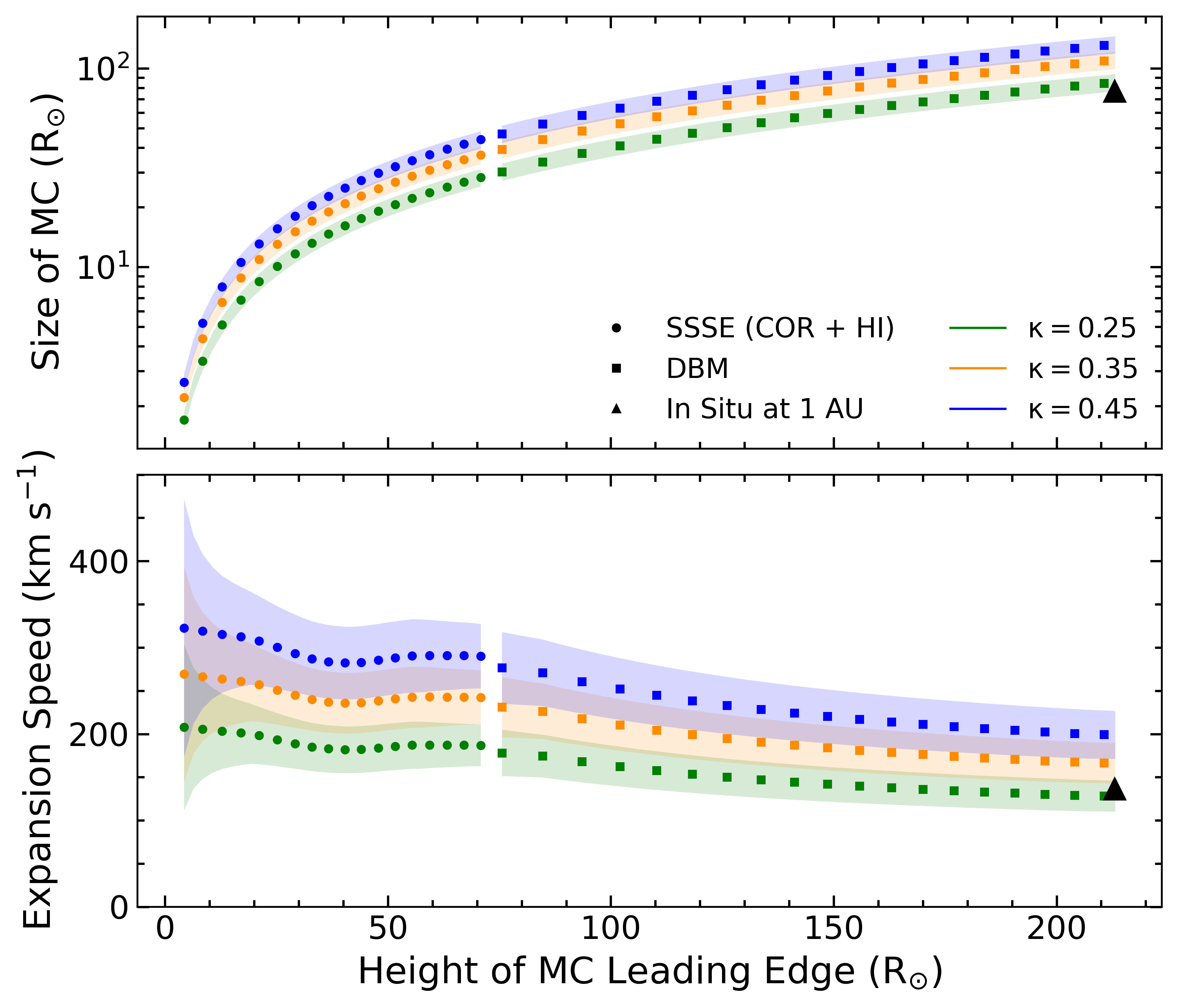}
    \caption{The upper and lower panels show the evolution of the radial size and expansion speed of the MC with the height of its LE for three different values of $\kappa$. The green, orange, and blue denote the $\kappa$ values as 0.25, 0.35, and 0.45, respectively. The filled circles and filled squares represent the estimates from the SSSE and DBM, respectively. The black-filled triangle denotes the in situ measured radial size and expansion speed at 1 AU from the conventional analysis approach to in situ observations of the CME/MC. The error bar over each data point is shown with transparent fill areas of the same color as the corresponding data point.}
    \label{fig:mat_kap}
\end{figure}

Our analysis finds differences in the CME/MC characteristics (arrival time, speeds, and radial size) between those derived from remote and directly measured from in situ observations. The difference is significantly large for the later segment (center and TE) of the MC. The radius of the MC at the arrival of LE ($R_1$) derived from 3D kinematics is 57.5 $R_\odot$, which is 1.5 times the radius (38.4 $R_\odot$) of the MC estimated from in situ measurements. This discrepancy could result from assuming a constant aspect ratio in analyzing remote observations of the CME during its interplanetary journey. This assumption implies that the rate of increase in the radius of the MC and the distance of its center is always the same in the IP medium. However, some studies suggest a change in the aspect ratio of the CME during its heliospheric journey \citep{Nieves-Chinchilla2018,Vršnak2019,Kay2021}. Therefore, we examine the size and expansion speed of the MC using different values of $\kappa$ for the same LE height of the CME as derived from SSSE+DBM (Figure~\ref{fig:mat_kap}). Although we lack the direct in situ measurements of MC size until it arrives at 1 AU, the estimate of the size at 1 AU and that from the GCS model applied on coronagraphic observations would enable us to note a change in the $\kappa$ for the CME as it evolves from near the Sun to near 1 AU.

In Figure~\ref{fig:mat_kap}, the upper and lower panels show the radial size and expansion speed of the CME for different values of $\kappa$ for its same LE height. The green, orange, and blue denote the three values of $\kappa$ as 0.25, 0.35, and 0.45, respectively. The black-filled triangle denotes the radial size of the MC and the expansion speed (from the conventional approach to in situ measurements) at the arrival of the LE at 1 AU. We note that as the value of $\kappa$ increases, the size of the CME/MC also increases. The value of $\kappa$ = 0.25 gives the size and expansion speed of the MC almost equal to that from the in situ measurements. Also, the in situ measured radius (38.4 $R_\odot$) of the MC on the arrival of its LE at 1 AU (213 $R_\odot$) provides $\kappa$ = 0.22, marking that the aspect ratio has decreased as the CME evolved from the Sun to 1 AU.

Although the aspect ratio derived from the GCS model ranges from 0.26 to 0.37, its value initially rises up to 0.37 at 6.5 $R_\odot$ and remains the same up to 13.7 $R_\odot$. This hints at the constancy of the aspect ratio beyond a certain height from the Sun \citep{Cremades2020}. However, the estimated value of $\kappa$ from in situ measurements at 1 AU suggests a decrease in $\kappa$ for the CME in the IP medium. Moreover, earlier studies have provided evidence of larger CME expansion closer to the Sun, which weakens significantly beyond 1 AU \citep{Bothmer1998,Liu2005,Leitner2007,Gulisano2010}. Therefore, we intuit a profile for a $\kappa$, which gives an initial rise to be constant up to a certain height, followed by a decrease with increasing height of the CME LE. From the anticipated profile of $\kappa$, we assume two profiles of $\kappa$, differing in terms of the height up to which $\kappa$ is constant and its further decline profile beyond that height. Corresponding to these two profiles of the $\kappa$, we estimate the size and expansion speed evolution of the CME.

\begin{figure}
    \centering
    \includegraphics[scale = 0.45]{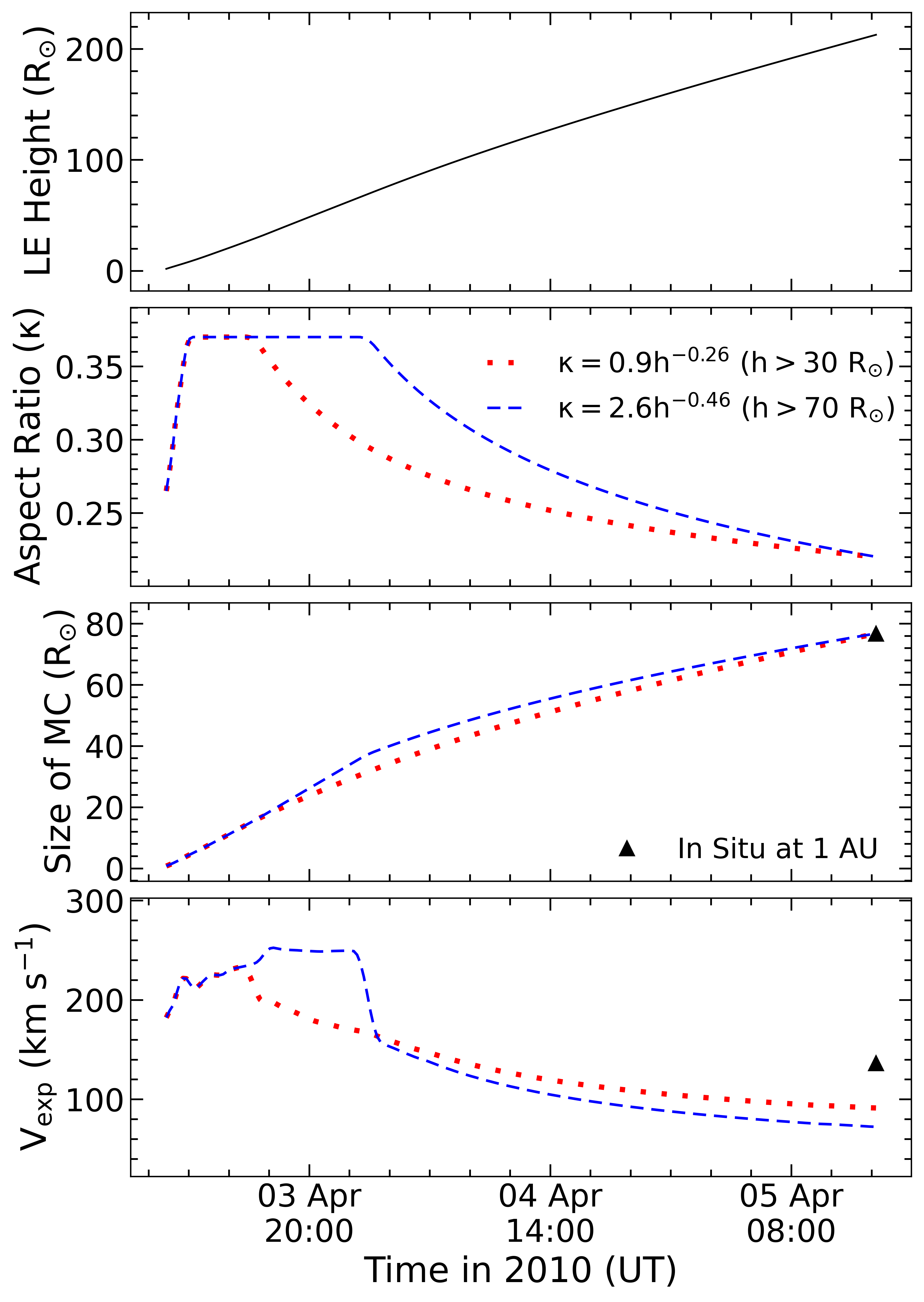}
    \caption{The top to bottom panels show the time evolution of the CME LE height, aspect ratio, radial size, and expansion speed. The red dotted and blue dashed lines denote the evolution of $\kappa$ in the IP medium, with two power laws decreasing the value of $\kappa$ for heights beyond 30 $R_\odot$ and 70 $R_\odot$, respectively. The black-filled triangle denotes the in situ measured radial size and expansion speed at 1 AU from the conventional analysis approach to in situ observations of the CME/MC.}
    \label{fig:kapp_evo}
\end{figure}

In Figure~\ref{fig:kapp_evo}, the panels from top to bottom show the 3D height of the CME LE, the aspect ratio for the same LE height, the size of the MC, and its expansion speed. The red and blue in the figure show the value of $\kappa$ constancy up to the height of the LE 30 $R_\odot$ and 70 $R_\odot$, respectively. Further, the two different power laws $\kappa = 0.9h^{-0.26}$ and $\kappa = 2.6h^{-0.46}$ shown with red and blue, respectively, decrease the value of $\kappa$ up to 0.22 after its constant value of 0.37. Using the conventional approach, the black-filled triangle denotes the in situ measured radial size and expansion speed of the MC at 1 AU. For the power law $\kappa = 2.6h^{-0.46}$, we notice a sharp decrease in the $\kappa$ evolution, resulting in a sharp decrease in the expansion speed. From Figure~\ref{fig:mat_kap} and ~\ref{fig:kapp_evo}, we note that the value of $\kappa$ as 0.25 at 1 AU perfectly matches the expansion speed of the MC from the conventional approach but not the in situ measured size, which corresponds to the value of $\kappa$ as 0.22. Such a disparity is possible due to inaccuracy in the expansion speed from the conventional approach to in situ observations \citep{Regnault2024}. This implies that the aspect ratio of CMEs plays an important role in governing the size and expansion speed of the MC during its IP evolution. Such an approach to evolve the aspect ratio of CMEs beyond coronal heights can be useful for comparing the size and expansion speeds of the CMEs from in situ measurements in the IP medium \citep{Zhuang2023}.

\section{Results and Discussion} \label{sec:resdis}

The present study focuses on a non-conventional approach to analyze in situ observations of the CMEs from single-point spacecraft. We selected the 2010 April 3 CME to estimate its radial size and instantaneous expansion speed at different instances (at the arrival of LE, center, and TE at 1 AU) during the passage of the MC over the in situ spacecraft. We also estimate the continuous evolution of the radial size and instantaneous expansion speed during the propagation of the CME from the Sun to 1 AU using the multi-point remote observations combined with the drag-based model. The independent estimates from conventional and non-conventional analysis approaches to in situ observations are compared with those from conventional 3D reconstruction methods to remote observations. Our analysis clearly demonstrates that despite continuously tracking the CME up to 1 AU in remote observations, the estimates of the arrival time of the CME substructures (LE, center, and TE), as well as the radial size and expansion speeds of the CME have large inconsistencies with those estimated from in situ observations. The inconsistency is especially significant when estimating the arrival time of rear-edge features (center and TE) following the CME LE. We also highlight that our non-conventional approach to analyzing in situ measurements could be more accurate than the conventional approach for deriving radial sizes and expansion speeds of CMEs at different instances.

The in situ measurements of the MC at 1 AU estimate its time duration as 23.6 hours, encompassing the LE and TE of the MC. We calculate the radial size of the MC at the arrival of LE as 76.8 $R_\odot$ by integrating the speed with time during the MC passage over the in situ spacecraft. We notice that the arrival of the size center of the MC at 1 AU and the time center of the MC duration are not synchronous. We further demonstrate a large impact of expansion on the asynchrony of time and size center by assuming a virtual speed profile with a steeper slope from LE to TE having identical boundaries as of the actual MC (in the left panel of Figure~\ref{fig:insitu_spe}). This suggests for an expanding MC; one needs to necessarily estimate its instantaneous expansion speed during its passage at in situ spacecraft \citep{Lugaz2020a}.

Our non-conventional approach to in situ observations, considering the non-constant expansion of the MC from its LE to TE, attempts to understand the evolution of the radial size (Figure~\ref{fig:MC_1Dcut}) during the passage of the MC at the single-point in situ spacecraft. In contrast to the conventional approach \citep{Owens2005,Zhuang2023}, we assume constant acceleration of the different features (LE, center, and TE) and estimate their propagation speeds at the same instance to further estimate the instantaneous expansion speed of the MC. The instantaneous expansion speeds at three successive instances ($t_1$, $t_2$, and $t_3$ in the in situ measurements) show the expected decreasing trend (in the third column of the bottom panel of Table~\ref{tab:tab_1}) and do not match the expansion speed derived from the conventional approach (136.5 km s$^{-1}$ in Section~\ref{sec:insitu}). We also deduce the distances traveled by each feature during two specific instances and the radial dimensions ($R_1$, $R_2$, and $R_3$) of the MC upon the arrival of distinct features (LE, center, and TE) at 1 AU. The radius $R_1$ of the MC derived from two different features, center and TE, are 37 $R_\odot$ and 36.45 $R_\odot$ (Table~\ref{tab:tab_2}), respectively. We note that the single-point in situ spacecraft can not directly provide $R_2$, and $R_3$ by measured speed and time, however, it can measure $R_1$. The estimates of $R_1$ from the non-conventional approach closely match the in situ measured radius as 38.4 $R_\odot$, underscoring the accuracy of the non-conventional approach.

We use the GCS model fit to contemporaneous coronagraphic observations of the CME from multiple viewpoints to investigate the kinematic evolution of the CME. The fitting parameters of the GCS model are consistent with earlier studies \citep{Möstl2010,Wood2011,Mishra2020} except for the tilt angle, which is difficult to constrain by manual fitting reliably. However, we note that the speed of the selected CME in the COR2 FOV differs significantly, with $\pm$200 km s$^{-1}$, among several earlier studies \citep{Möstl2010,Wood2011,Xie2012,Mishra2020}, which is most probably due to differing methods to calculate the speeds from the height of the CME LE. We also demonstrate the considerable effect of different fitting techniques on deriving the speed, especially at the end points of the height-time profile (Figure~\ref{fig:diff_tt}). Moving box linear fit for speed estimation from heights avoids systematic unphysical fluctuations and keeps the real short-term variations in the CME speed. We also demonstrate that uncertainties of $\pm$100 km s$^{-1}$ in the speed, and assuming it to be constant beyond COR2, can give an error of 10-20 hours in estimating arrival time for a typical fast and slow speed CME at 1 AU. Therefore, we further tracked the CME in HI observations and estimated the 3D kinematics to be used to derive the CME parameters at 1 AU.

To investigate the evolution of the CME beyond COR2, the SSSE method is used on HI observations. The earlier studies have only estimated the evolution of CME LE in contrast to the present study, which also examines the evolution of the center and TE of the CME. The tracking of TE of the CME in remote observations is difficult and contains uncertainties \citep{DeForest2011,Mishra2015}; therefore, we use the aspect ratio of the CME to estimate the height-time evolution of features other than the CME LE \citep{Zhuang2023}. This implies that, in our study, the CME center and TE kinematics depend solely on the aspect ratio obtained from the GCS fitting in the COR FOV. Moreover, unlike the earlier studies, which often use single-spacecraft observations or use the geometry of the CME from an ad-hoc assumption \citep{Möstl2010,Wood2011,Liu2011,Xie2012,Mishra2013,Colaninno2013,Mishra2014a, Xiaolei2021}, we implement SSSE taking inputs of the angular width from the GCS model while assuming it to be constant in the HI FOV. To investigate the evolution of the CME at distances beyond the final height of the CME LE derived from the SSSE, we use the DBM \citep{Vršnak2013} for the CME by estimating the actual drag parameter of the CME in contrast to earlier attempts that adopted a statistical range of the drag parameter \citep{Mishra2013,Mishra2014a}. Earlier studies have also used GCS model inputs to HI-based reconstruction methods and further estimates from the HI-based methods as inputs to the DBM \citep{Rollett2016,Amerstorfer2018} to predict the CME arrival time and speed at the Earth. In our study, the time evolution of the radial sizes and expansion speeds jointly from GCS+SSSE+DBM enabled us to compare these estimates to those derived from in situ measurements at 1 AU.

The arrival time of different features/substructures (especially center and TE) from remote observations (GCS+SSSE) combined with DBM significantly differs from those obtained from in situ observations. The estimated propagation speed of different features (LE, center, and TE) at the same instance from remote observations substantially differ from measured in situ observations as well as the estimation from our non-conventional approach to in situ observations (as shown in the third panel of the Table~\ref{tab:tab_1}). We also compare the estimate of $R_1$ and find that its value from GCS+SSSE+DBM is 1.5 times the actual in situ measured radius. We note that the traveled distances by each feature derived from GCS+SSSE+DBM methods applied to remote observations are significantly greater than those derived from in situ observations (Table~\ref{tab:tab_2}). The inconsistency between estimates from remote and in situ observations may come from assuming a constant aspect ratio for CME during its continuous journey much away from the Sun \citep{Savani2011a,Kay2021}, which creates larger uncertainties in the derived height of the center and TE of the CME.

We examine the impact of the distance-dependent evolution in the GCS model-derived extrapolated aspect ratio on the radial size and expansion speed of the MC. The value of $\kappa$ from the in situ measured radial size at 1 AU is 0.22, which is smaller than the value of 0.37 from the GCS model in the corona. This indicates that the $\kappa$ value possibly decreased at farther distances from the Sun. Therefore, the expansion behavior of the CME in the corona may not be consistent with that in the IP medium \citep{Savani2011,Zhuang2023}. Our assumed profile of $\kappa$ with two different power laws, based on matching the near-Sun and 1 AU estimates of $\kappa$, indicates the possibility of different evolution of the expansion speeds of the CMEs. The variations in the aspect ratio and radial size of the CMEs, noted from remote observations, could have been directly verified by employing the in situ observations from multiple radially aligned spacecraft.

In our study, the derived radial sizes and instantaneous expansion speeds depend considerably on the constant acceleration of the CME features used in our non-conventional analysis approach. Additionally, the acceleration estimation from the in situ speed-time profile has uncertainties that could be minimized using multiple in situ spacecraft measuring the speeds of the same feature at different instances. Although our study utilized multi-point remote observations, our analysis relied solely on single-point in situ measurements to associate the estimates from both sets of observations. To further validate the efficacy and superior performance of our non-conventional approach over the conventional approach in analyzing in situ observations of CMEs, future investigations should capitalize on multiple radially aligned in situ spacecraft observations \citep{Good2016,Davies2021}.

Our study ignores observational bias due to the geometric selection effect caused by the in situ spacecraft sampling path through a large CME geometry \citep{Zhang2013}; therefore, comparing global measurements from remote observations with local measurements from in situ may introduce additional inconsistencies. Future studies can focus on estimating the acceleration of different features of CMEs to be used in our non-conventional approach. Also, the instantaneous expansion speed could be obtained by sampling CME LE with one in situ spacecraft and simultaneously sampling the TE or center of the MC with another in situ spacecraft \citep{Regnault2024}. To enhance the reliability of our findings that CMEs experience changes in the expansion speeds during their complete passage to the in situ spacecraft, it is imperative to use our non-conventional analysis approach to in situ observations and analyze a broader spectrum of CMEs/MCs exhibiting varying speeds and accelerations.

Our study clearly shows the difficulty in comparing remote to in situ observations of different CME features despite choosing the 2010 April 3 CME that should have been an ideal candidate for such a comparison. This is because the chosen CME shows minimal change in the dynamics beyond coronagraphic heights, and its different features are well-identified in the in situ measurements at 1 AU. This perhaps highlights the best-achieved performance of the state-of-the-art approaches to analyzing remote and in situ observations in forecasting the radial size, expansion speeds, and impact duration of CMEs at 1 AU. We highlight the major gap in our comprehension of the radial sizes and expansion behaviors of CMEs, predominantly attributable to the constraints of single-point in situ observations and the conventional approach to analyzing them. Future investigations in this realm, using the introduced non-conventional approach to multiple in situ spacecraft observing at unprecedented distances close to the Sun \citep{Fox2016,Muller2020}, would help understand physical processes dictating the radial sizes and expansion of CMEs during their dynamic interaction with the ambient solar wind.

\section*{Acknowledgements}

We thank the STEREO (COR and HI), SOHO (LASCO C2 and C3), and Wind spacecraft team members for making their observations publicly available. We also thank the anonymous referee for their careful review and insightful comments that helped us to improve the manuscript.

\section*{Data Availability}

The data sets used and generated for this study are available upon request from the corresponding author.


\bsp	
\label{lastpage}
\end{document}